\documentclass{article} 

\usepackage[english]{babel} 
\usepackage{amssymb}
\usepackage{amsmath}
\usepackage{physics}
\usepackage{txfonts}
\usepackage{geometry}[2cm,2cm,2cm,2cm]
\usepackage[version=4]{mhchem}
\usepackage{mathdots}
\usepackage[classicReIm]{kpfonts}
\usepackage{siunitx}
\usepackage[dvips]{graphicx} 

\begin{document}

\begin{center}
\Large 
\textbf{Modelling functional properties of ferroelectric oxide thin films}
\textbf{with a three-domain structure}
\end{center}	
\normalsize
\vspace{2em}

\noindent \textbf{E.P. Houwman$^1$, K. Vergeer$^{1,2}$, G. Koster$^1$ and G. Rijnders$^1$ }

\vspace{2em}

\noindent $^1$ Inorganic Material Science, MESA+ Institute of Nanotechnology, University of Twente, \\ Enschede, The Netherlands

\noindent $^2$ Materials innovation institute (M2i), Delft, The Netherlands

\noindent Corresponding author: e.p.houwman@utwente.nl

\vspace{2em}
\textbf{Abstract }
\vspace{1em}

The properties of a ferroelectric, (001)-oriented, thin film clamped to a substrate are investigated analytically and numerically. The emphasis is on the tetragonal, polydomain, ferroelectric phase, using a three domain structure, as is observed experimentally. The previously used, very restrictive set of boundary conditions, arising from the domain walls, is relaxed, creating more modes for energy relaxation. It is argued that this approach gives a more realistic description of the clamped ferroelectric film.

It is shown that for the ferroelectric oxides \ce{PbZr_{1-x}Ti_xO_3} the tetragonal, polydomain phase is present over a wide range of substrate induced strains for \ce{x_{Ti}}$\ge$0.5, corresponding to the tetragonal side of the bulk phase diagram. A polydomain, rhombohedral phase is present for \ce{x_{Ti}}$<$0.5, at the bulk rhombohedral side. Phase-temperature diagrams, and ferroelectric, dielectric and piezoelectric properties, as well as lattice parameters, are calculated as function of substrate induced strain and applied field. The analytical formulation allows the decomposition of these properties into three different causes: domain wall motion, field induced elastic effects and piezoelectric effects. It is found that domain wall motion and polarization rotation of the in-plane oriented domains under an applied field contribute most to the properties, while the out-of-plane oriented domains hardly contribute.

\vspace{2em}
\textbf{1 Introduction}
\vspace{1em}
\begin{sloppypar}
Perovskite ferroelectric and piezoelectric materials and notably the solid solution
group 
\ce{PbZr_{1-x}Ti_xO_3} (or short PZT), which shows the strongest ferro/ piezoelectric 
properties up to relatively high temperatures is of great interest for many applications \cite{Lines1977,Scott2007,Davies2008}. 
In thin film form these materials have been investigated extensively for use in 
FeRAM memory applications and piezoelectric driven devices. 
In bulk the most used composition is that of the Morphotropic Phase Boundary (MPB) for \ce{x_{Ti}} $\approx$ 0.48, between the tetragonal phase (\ce{x_{Ti}} $>$ 0.48) and the rhombohedral phase (\ce{x_{Ti}} $<$ 0.48) at room temperature. 
Despite the long history of research into ferro/piezoelectric materials and the thorough understanding of the basic mechanisms in these materials it is in practice hard to describe the properties of PZT (or any other ferro/piezoelectric material) in real devices quantitatively very accurately, due to the large number of extrinsic effects that can play a role. 
Therefore there is a need for models that can separate intrinsic and extrinsic contributions to the various functional properties, in realistic thin films.
\end{sloppypar}
Ferro/piezoelectric thin films are of great importance for future MEMS applications because of their promise of increased miniaturization of mechanical devices. 
Integration of the perovskite crystal structure with different electrode and substrates (especially with Si) is therefore of utmost importance. There are many design variables that influence the properties of the thin film, such as a) the substrate material and its crystalline structure - single or poly-crystalline or even amorphous (glass) -- and orientation, determines to a large extent the film growth. b) Additional buffer layers between the substrate and the ferroelectric film may alleviate possible lattice mismatches \cite{Dekkers2009}. Oxide nanosheets can even allow coherent growth on amorphous substrates \cite{Bayraktar2014}. c) The bottom electrode material also acts as a buffer layer and is likely to determine to a large extend the initial growth layer of the functional ferroelectric layer. The interaction of the ferroelectric film with the electrode layers can cause interface related electrical effects such as voltage self-bias of the device and changes in the polarization hysteresis loop due to an interface passive layer \cite{Tagantsev2006}, as well as domain wall pinning affecting the coercive field \cite{Tagantsev2004}. d) The substrate/buffer layer stack in most cases dictates the growth orientation of the functional layer \cite{Dekkers2009}. If the polarization axis is strongly coupled to the crystal structure, as is the case for compositions away from the MPB, this may in turn affect device properties. e) As one would expect, the higher the in-plane alignment of the crystal structure of the individual grains in the film, the less effect of the grain boundaries and the highest ferroelectric properties on the film properties were observed, even without electrical poling, as was recently shown in a comparative study \cite{Houwman}]. Further such devices appear not to suffer from aging effects, a property that is very important for many applications. f) When aiming for the largest piezoelectric coefficients experimentalists tend to choose the MPB composition., However it is not self-evident to assume that bulk properties apply in thin film conditions. For example for (110)-oriented PZT thin films on Si the highest piezo-electric coefficient $e_{31}$ was found for $x \approx 0.6$ and not for 0.48 \cite{Wan}. g) The ferroelectric domain structure plays a very important role in the properties of ferro and piezoelectrics \cite{Tagantsev2010}. For bulk single crystal devices this is well understood and one often engineers the domain structure to optimize device properties by using specific crystal cuts and polarization directions. However, in the case of thin film devices the importance of the domain structure and its role in device properties is well recognized, but the theoretical description is still a field in development. With the increasing control over thin film growth one may also envision the design of domain structures in thin film devices.
\vspace{1em}
With the advent of high quality ferroelectric/piezoelectric thin films there has been much theoretical development in understanding many of the extrinsic effects, recently summarized in the review book of Tagantsev \textit{et al.} [10]. Many of these effects are related to the crystalline quality of the grown films. Experimentally it is not easy to separate extrinsic effects from the intrinsic properties of the thin film, that are modified by the film clamping. However, in well-defined situations it is possible to model the modified properties of clamped ferro/piezoelectric thin films. Thus if such a model is available one can make a better effort to separate other extrinsic effects, arising from for example grain boundaries, from that of the film clamping. This is of great importance for understanding thin film properties and the improvement of thin films for demanding applications.

There is already a significant number of theoretical papers describing the 
properties of clamped, epitaxial ferroelectric/piezoelectric thin films,
also considering polydomain phases of the tetragonal compositions. 
Most literature considers the case of (001)-oriented epitaxial thin films. In a series of papers of Pertsev, Zembilgotov, Kukhar and coworkers \cite{Perstev1995,Pertsev1996,Pertsev1998,Pertsev1999,Pertsev2000,Koukhar2001,Pertsev2003,Kukhar2006} the polydomain structure of the tetragonal phase has been modelled in great detail. 
In this description it is (somewhat implicitely) assumed that the considered tetragonal $c/a$ domain structure is representative for a film in which $c/a$, $c/b$ and $a/b$ structures are present.
$a$ and $b$ are domains with orthogonal polarization directions in the film plane, the $c$ domain has the polarization oriented out-of-plane). 
Considering only a $c/a$ domain structure imposes strong strain boundary conditions in one direction, that lead to divergences in film properties. Here we reconsider the problem by taking the experimentally observed three-domain structure into account. 
Although it has been recognized that in thick films the domain structure of a (001)-oriented film is constituted of both $c/a$ and $c/b$ domain substructures \cite{Roiburd1976,Roytburd2001}, to our knowledge there is no detailed descriptive model of a film with such a polydomain structure. 
Recently Ouyang \textit{et al.} gave a thermodynamic analysis of a clamped ferroelectric  (001)-oriented film, using a linearized model \cite{Ouyang}. 
However there neither the coupling of polarization with the stress (piezoelectric effect) nor the rotation of the polarization in in-plane domains under the influence of an electric field normal to the film were considered. 

In this paper we give a non-linear thermodynamic analysis of the polydomain, clamped,  (001)-oriented, epitaxial film, taking into account the possibility of polarization rotation. We argue that the contribution of domain walls and micro-stresses to the total energy can be neglected under certain conditions. 
It is shown that polarization rotation plays a major role in the film properties in the three-domain, tetragonal phase, especially close to the MPB, and should therefore not be neglected in the analysis. 
To this end we modified an earlier model in literature for describing the ferro/piezoelectric properties of a symmetrically clamped, ferroelectric, polydomain thick film \cite{Kukhar2006} by reformulating the approach with a two-domain $c/a$ structure to effectively a three-dimensional $c/a/b$ domain structure and investigate the consequences for the film properties. For thick films it is assumed that the properties are homogenous in the third out-of-plane dimension, therefore the model is limited to thick PZT thin films of at least a few hundred nm. From an application point of view these are thicknesses that are used in many Si-based MEMS devices. Further the model assumes (001)-oriented, epitaxial films without grain boundaries. A second difference with the earlier model is that we argue that the strict boundary conditions imposed previously on the $c/a$ domain walls may be relaxed. In fact only global boundary conditions for the film are assumed and the domain walls, connecting domains with homogeneous properties (which are affected only by the global boundary conditions), are treated as small volume planes in which all strain and polarization gradients are concentrated. This approach allows us to obtain analytical expressions for the strain and applied field dependent properties of the three-domain phase. Temperature-strain and applied field-strain phase diagrams of the thin film are calculated as function of composition, as well as properties as piezoelectric and dielectric coefficients and lattice parameters as function of substrate induced strain and applied field. In contrast to the one-dimensional approach only tetragonal and rhombohedral polydomain and single domain phases are found, but no additional intermediate phases arise. For compositions close to the MPB the previously found rhombohedral phase and intermediate phases are replaced by the two-dimensional tetragonal polydomain phase. We expect that the model is equally well applicable to other materials and can be modified in the future to other film orientations and more disordered films.

\vspace{2em}
\textbf{2  Polydomain ferroelectric thin films}
\vspace{1em}

To understand the relation between film properties and the above mentioned structural variables models are needed that describe the dielectric, ferroelectric and piezoelectric behavior of thin films. Since in most devices the thin film device area is much larger than typical ferroelectric domain sizes such models should take the domain structure into account, as well as its response to external forces, such as applied stress or electrical field. Ferroelectric domain formation driven by elastic constraints has been studied by various authors. Tagantsev \textit{et al.} \cite{Tagantsev2010} summarized recently the various models existing in literature. Three principal approaches can be distinguished. In the \textit{mean-strain approach}, initiated by Roitburd \cite{Roiburd1976,Roytburd2001} and more recently extended to a description of the three-domain architecture \cite{Ouyang}, the average mechanical energy of the system is minimized by the creation of subdomains, without considering the coupling to polarization. In the second approach, based on the \textit{Landau-Devonshire theory of the dense domain structure} \cite{Koukhar2001,Kukhar2006} one takes also the stress dependence of the order parameter (the polarization) by the piezoelectric effect into account. In the numerical \textit{phase-field approach} \cite{Li2001,Li2002} the polarization relaxes using time-dependent Ginzburg-Landau equations. 

\begin{figure}
	\includegraphics[width=0.6\linewidth]{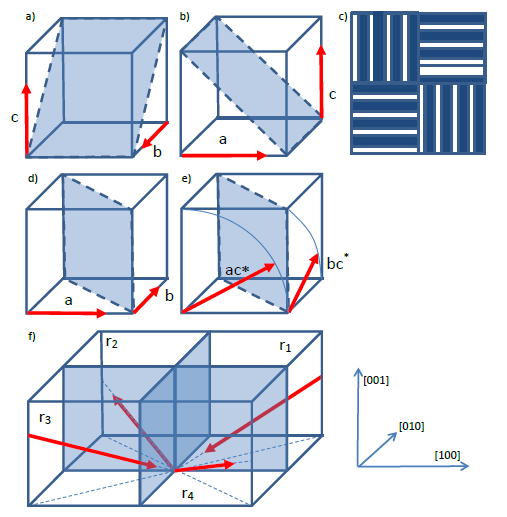}
	\includegraphics[width=0.4\linewidth]{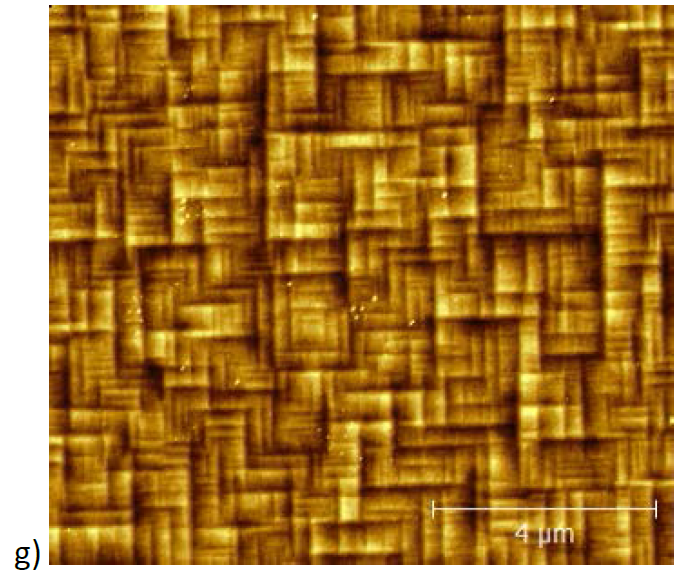}
	\caption[domain structures]{Schematic $c/b$ (a) and $c/a$ (b) domain substructures. (c) Top view of $c/a/b$ domain structure. (d) $a/b$ domain structure in zero and (e) finite field. (f) $r$-phase structure in zero field. (g) Atomic Force Measurement of the surface of a PZT(\ce{x_{Ti}}$=0.6$) thin film revealing the cellular domain structure, schematically drawn in (c); shown area is 3x3 $\mu m^2$ }
	\label{domainstructures}
\end{figure}

Here we reconsider the model of Koukhar \textit{et al}. \cite{Koukhar2001,Kukhar2006} for several reasons. There a two-domain structure is considered in a one-dimensional approach, whereas experimentally in tetragonal PZT thin films (with a thickness of at least a few \SI{100}{\nano\meter}) generally a three-domain architecture is observed, i.e. the (001)-oriented tetragonal films not only show $c/a$, but also $c/b$  and $a/b$ -subdomain structures (see Fig.\ref{domainstructures}g) are visible. Further we will argue that the very strict boundary conditions imposed by Koukhar and Pertsev can be relaxed. In that aspect we follow the approach of Roytburd who considers only macroscopic (global) boundary conditions. Taking these two differences into account we model the consequences for the film properties analytically as well as numerically. 

Engineering oriented experimentalists measure film properties as function of applied field rather than of temperature. Therefore we also analyze the film properties as function of applied field. We note that the model is in essence static and does not account for an eventual frequency dependence of the domain wall motion, which may arise from the coupling of domain walls with defects (domain wall pinning). The analytic expressions obtained allows one to distinguish quantitatively the contribution from extrinsic contributions such as domain wall motion and stress from intrinsic contributions to the effective film properties. Although not pursued further here this separation may also allow one to make qualitative statements on the frequency dependence of film properties.

The paper is structured as follows. Paragraph 3 describes a clamped thin film and discusses under which conditions the description can be simplified to obtain a mathematically treatable problem, without losing the essential characteristics of a realistic, clamped thin film. A general expression for the free energy of a polydomain, clamped thin film and the boundary conditions is given in paragraph 3.1. The case of a thin film with a tetragonal PZT composition in the polydomain phase is treated in paragraph 3.2, while the properties of this phase are described in the paragraphs 3.3-3.5. Results of the numerical analysis of all phases are presented and their interpretation in terms of the derived analytical descriptions are given in paragraph 4. In paragraph 5 we summarize the main results. The analytical approach concentrates on the polydomain, tetragonal phase, because that is the phase one mostly encounters in experimental work. However in the numerical analysis also other phases, such as the polydomain tetragonal $a_1/a_2$-phase, that lacks out-of-plane oriented tetragonal domains, the polydomain rhombohedral $r$- phase and the mono-domain $c$ - phase are described. The mathematical description of the latter phases can be found piecewise in several publications in more or less detail. In the Supplemental Material we summarize these results relevant to this paper, making some extensions not described in literature so far.

\vspace{2em}
\textbf{3 Model of a polydomain PZT thin film} \newline

We consider thick, single-crystalline, epitaxial films (here specifically made of PZT) that are grown in the paraelectric state at deposition temperature $T_d$ on a thick substrate of a dissimilar material (for example perovskites like \ce{SrTiO3} (\ce{STO}), \ce{DyScO3} (\ce{DSO}) and others, as well as Si, with appropriate buffer layers). If the film thickness is larger than a few \SI{100} {\nano\meter} the bulk of the film can generally be considered to be fully relaxed at $T_d$ due to the incorporation of growth defects in the initial growth layer during deposition, resolving the epitaxial lattice mismatch between film and substrate (or eventual buffer or bottom electrode layer). Since $T_d$ is generally above the paraelectric-ferroelectric transition temperature $T_C$ the PZT film is in its cubic parent phase during deposition. The initial growth layer has a thickness of the order of at most a few 10's of nm. This is of the order of a few percent of the thickness range of films normally used in piezoelectric device applications, 500-\SI{2000}{\nano\meter}. Upon cool-down tensile or compressive stress builds up in the film due to the difference in thermal expansion coefficients between film and substrate. Under the so-called clamped film condition it is assumed that the thick substrate does not deform. 

Here and in literature the misfit strain is defined as

\begin{equation}
\label{Sm0}
S^0_m(T) = \left( \frac{a^*_s-a_0}{a_0} \right)_T
\end{equation}

The in-plane lattice parameter of the clamped film is equal to the effective substrate parameter at temperature $T$, $a^*_s$, while $a_0$ is the equivalent cubic lattice parameter of the paraelectric phase of the stress-free film material at temperature $T$. The misfit strain is due to the thermal mismatch with the substrate and the paraelectric-ferroelectric phase transition. In appendix A the misfit strain is expressed in terms of the thermal expansion coefficients and the phase transition and its value is calculated for different substrate-PZT(\ce{x_{Ti}}) combinations. When cooling the film through the Curie temperature $T_C$ the thermally induced stress can in some cases be (partly) resolved by the formation of a ferroelastic domain structure. This is for example possible for the well-known $c/a$ domain structure of a film with a tetragonal PZT composition. For a $(001)$-oriented film this structure consists of alternating 45\textdegree inclined slabs of $c$ and $a$ domains. Another strain relaxation mechanism is rotation of the polarization vector in the domains, which changes the unit cell lattice parameters through the coupling between strain and polarization. Koukhar \textit{et al} \cite{Koukhar2001} argued that in a relatively thick film the domain width $D$ is much smaller than the film thickness $H$, so that the polarization and strain fields within each domain can be considered to be homogeneous. Therefore also the energy density in a polydomain epitaxial film is piecewise homogeneous. Further the energy contribution to the total free energy by the thin layers (with thickness $h\ (\sim D)\ll H$) with inhomogeneous internal fields near the top and bottom interface can be neglected. Further it was argued that the energy contribution of the domain wall self-energy is small under the condition $D\ll H$, which holds for thick films \cite{Koukhar2001,Kukhar2006} and can therefore be neglected. The latter condition can be relaxed assuming that the domain wall energy per unit domain wall area is constant and that the total area of the domain walls does not change (thus that no domain walls are created or disappear). This is the case when the domain wall positions only shift under varying mechanical or electrical field conditions. In that case the total domain wall energy per unit volume does not change and is just an additional, but constant energy term in the expression for the total free energy of the film. Further we consider a device structure with short-circuited or voltage-biased top and bottom electrodes and no internal charges or screening, thus depolarization does not affect the total energy. 

Fig.\ref{domainstructures} shows schematically the possible domain structures considered in this paper. The domains and polarization vectors are defined in terms of the pseudocubic representation of the (001)-oriented PZT unit cell. The approach to obtain a description of the polydomain state, as is presented here, is different from that followed in literature in three aspects. These differences give rise to significant qualitative and quantitative differences in the properties of the polydomain phases. 

1) Several polydomain phases are possible in the model, however there is only one phase that gives rise to an additional parameter, the domain volume fraction of $c$ domains, $\phi =V_c/V_{tot}$, in the polydomain tetragonal phase, which can be used to minimize the total energy. In the one-dimensional problem description the $c/a$ domain structure is assumed to be representative for the poly-domain state of a two-dimensional film. The consequence of this assumption is that the strain and stress states of the domains are \textit{a}symmetric in the two in-plane directions \cite{Kukhar2006}. \textit{Here}, the presence of both $c/a$ and $c/b\ $subdomain structures are considered, which form together the $c/b/a$ phase (Figs.\ref{domainstructures}a,b,c). Under large tensile strains one expects that all polarization vectors lie in the film plane, resulting in a two-domain $a/b$ domain structure (Fig.\ref{domainstructures}d), with \{110\} domain walls. When an out-of-plane electrical field is applied to the $a/b$ phase, the polarization vectors tilt slightly out of the plane to give the $ac^*/bc^*$ domain structure (Fig.\ref{domainstructures}e), causing a slight rotation of the domain wall. (With $x^*$ we will denote a small rotation (component) of the polarization vector in the $x$-direction). For small $c^*$ components the $ac^*$ and $bc^*$domains can also be combined with $c$-domains into $c/ac^*$ and $c/bc^*$ subdomain structures $\ $similar to the $c/a$ and $c/b\ $domains, to create the $c/bc^*/ac^*$ phase. Again this is accompanied by a small rotation of the domain walls. Although the basic crystal structure may be for example tetragonal, under the influence of stress or electrical fields the crystal symmetry can change and the polarization orientation changes accordingly. The rhombohedral polydomain structure $r_1/r_2/r_3/r_4$ (in short the $r$-phase) (Fig.\ref{domainstructures}f) consists of equal fractions of rhombohedral domains with the polarization vector in the (110) planes. Of course this phase is expected to arise in the case of rhombohedral compositions, but may also be the consequence of strain. For example in the one-dimensional approach it was found to be the lowest energy state in a certain strain range for compositions near the MPB \cite{Kukhar2006}. The phase change from $c/b/a$ to the $a/b\ $or $c$ phase goes gradually with a gradual change of the domain wall structure, but the change from the $c/b/a$ phase to the $r$-phase requires a significant rearrangement of domain walls and there may be an energy barrier to do so. The present model does not contain such energy barriers and can therefore not describe $c/b/a$ to $r$-phase (or vice versa) phase transitions under changing strain or applied field conditions, but only produces the minimum energy configuration for given strain and field.

2) In the one-dimensional-model fairly `strict' microscopic mechanical boundary conditions are assumed to be present at the domain walls (details below). This imposes strong restrictions on the stresses and strains in the domains. \textit{Here}, we assume `relaxed' mechanical boundary conditions. We argue that this is justified if one considers the domain walls to be regions of finite thickness in which all the stress, strain and polarization gradients are confined, while in the domains no gradients are present. Since the (change of the) energy of the domain walls can be neglected, it follows that the film consists of domains in which the stress and strain fields are piecewise homogeneous and which are coupled by the macroscopic boundary conditions only. These impose dimensional constraints on the film, which can also be interpreted as the requirements that the domains remain connected with each other and the substrate (the latter is the clamping condition). These assumptions also allow for the possibility of different stress and strain states to exist in the domains at both sides of a domain wall. This was not possible in the one-dimensional approach. Neglecting the local mechanical coupling between domains at both sides of a domain wall and with the substrate poses a simplification, which causes more stress relaxation than is possible under the more strict mechanical boundary conditions.

3) The electrical boundary conditions applicable at the domain wall relate the polarization orientations and polarization values on both sides of the domain wall. In the one-dimensional approach fixed angles between the orientations were assumed, specifically an uncharged, 90\textdegree domain wall in the $c/a$ domain phase. Consequently the polarization vectors in both domains are orthogonal. It is well known that the angle $\alpha$ between the polarization vectors depends on the short and long tetragonal lattice parameters $a_T$ and $c_T$ as $\alpha =2atan(a_T/c_T)$ \cite{Kittel}. Leaving the domain wall angle $\alpha $ free makes it possible that the polarization vector length and orientation in differently oriented domains may vary under varying stress conditions due to an applied external electrical field, thus allowing polarization rotation in the model. We assume that the domain wall angle $\alpha $ and orientation in the film adapts to minimize the domain wall energy  (keeping the domain wall uncharged), when the polarization at one or both sides of the wall rotates under the influence of electrical field or stress. 

Summarizing, the only condition imposed on the domain walls is that they are formed in such a way that the total energy of the film is minimized, but the contribution of a change in the domain wall energy to the total energy can be neglected. This condition is assumed to be applicable to all types of domain walls, thus not only to those between $c$ and $a$ domains, respectively $c$ and $b$ domains and $a$ and $b$ domains , but also those separating $c/a$, $a/b$ and $c/b$ domain structures.

\vspace{2em}
\textbf{3.1 Free energy of a polydomain, clamped $\mathbf{(001)}$-oriented thin film}	
\vspace{1em}

The Gibbs energy of a bulk PZT ferroelectric is usually given by a sixth order polynomial in the polarization components $P_i$ and the internal mechanical stresses $\sigma_{ij}$ \cite{Haun-1},
	
\begin{equation}
\label{eqG}
	\begin{aligned}
	G 	=	&G_0-\frac{1}{2}s_{klmn}\sigma_{kl}\sigma_{mn}-Q_{klmn}\sigma_{kl}P_mP_n \\ 
	G_0 = &\alpha_1(P^2_1+P^2_2+P^2_3)+ \alpha_{11}(P^4_1+P^4_2+P^4_3)+\alpha_{111}(P^6_1+P^6_2+P^6_3) \\
	&+ \alpha_{12}(P^2_1P^2_2+P^2_3P^2_2+P^2_1P^2_3)+\alpha_{123}(P^2_1P^2_2P^2_3) \\
	&+ \alpha_{112} \left(P^4_1(P^2_2+P^2_3)+P^4_2(P^2_1+P^2_3)+P^4_3(P^2_2+P^2_1) \right) 
	\end{aligned}
\end{equation}
	
$\alpha_1,\alpha_{kl},\alpha_{klm}$ are the dielectric and higher order stiffness coefficients at constant stress. Values for these parameters are given in \cite{Haun-2}. The temperature dependence of the properties are determined by that of the only temperature dependent parameter in the model, $\alpha_1=(T-T_C)/2\varepsilon_0C$, where $T_C$ , $C$ and $\varepsilon_0$ are the bulk Curie-Weiss temperature, Curie-Weiss constant and the vacuum permittivity respectively. $s_{klmn}=s_{ij}$ are the elastic compliances at constant polarization and $Q_{klmn}=Q_{ij}$ the electrostrictive constants. Values for the compliances are given in \cite{Pertsev2003} and for the electrostrictive constants in \cite{Haun-2}. The subindices $i$ and $j$ are used for the Voigt notation, which is used from here on. From the thermodynamic relations $S_i=-\partial G \ \partial \sigma_i$ the strains $S_i$ can be obtained. For a clamped (001)-oriented thin film with an applied electrical field $\vec{E}=(0,0,E)$ between the top and bottom electrode the appropriate thermodynamic potential is the Helmholtz free-energy $F=G+\sum^6_{i=1}{S_i}\sigma_i-EP_3$ \cite{Koukhar2001}. Eliminating the strains in $F$ with help of the thermodynamic relations one obtains for domain $x$ the free energy density  
	
\begin{equation}
\label{eqFx}
	\begin{aligned}
	F_x = &\alpha_1(P^2_{x1}+P^2_{x2}+P^2_{x3})+\alpha_{11}(P^4_{x1}+P^4_{x2}+P^4_{x3})+
	\alpha_{111}(P^6_{x1}+P^6_{x2}+P^6_{x3}) \\ 
	&+\alpha_{12}(P^2_{x1}P^2_{x2}+P^2_{x3}P^2_{x2}+P^2_{x1}P^2_{x3})+\alpha _{123}(P^2_{x1}P^2_{x2}P^2_{x3})  \\ 
	&+\alpha_{112} \left(P^4_{x1}(P^2_{x2}+P^2_{x3})+P^4_{x2}(P^2_{x1}+P^2_{x3})+P^4_{x3}(P^2_{x2}+P^2_{x1})\right) \\
	&+ \frac{s_{11}}{2}(\sigma^2_{x1}+\sigma^2_{x2}+\sigma ^2_{x3})+s_{12}(\sigma_{x1}\sigma _{x2}+\sigma_{x1}\sigma_{x3}+\sigma_{x3}\sigma_{x2}) \\ 
	&+ \frac{s_{44}}{2}(\sigma^2_{x4}+\sigma^2_{x5}+\sigma^2_{x6})-EP_{x3}	
	\end{aligned}
\end{equation}

We will use the index $x$ to denote the domain type ($x=a,b,c,r$). Within the domain the properties are homogeneous. For the $c$, $a\ $and $b$ domain the long axis is respectively in the pseudocubic (001) out-of-plane direction (subindex 3 of the polarization vector), the (100) in-plane (subindex 1) and the (010) in-plane (subindex 2) directions. The total free energy of the film is the sum of the energy contributions of the different domains and the total domain wall energy
	
\begin{equation}
	\label{eqFav}
	<F> = \sum_x{\phi_xF_x(P_{xi},\sigma_{xi},E)} + F_{DW}
\end{equation}

Here ${\phi }_x$ is the domain fraction of domain type $x$. The clamped substrate condition imposes macroscopic mechanical boundary conditions in both in-plane directions

\begin{subequations}
\label{eqS126BC}
\begin{align}
		S^0_m = <S_1> &= \sum_x{\phi_xS_{x1}} \\
		S^0_m = <S_2> &= \sum_x{\phi_xS_{x2}} \\
		<S_6> &= \sum_x{\phi_xS_{x6}} = 0
\end{align}
		
The first two conditions describe the coupling of the film to the substrate. The last condition implies that there is no net shear in the film plane. Further macroscopically there are no net forces acting on the upper surface, hence the corresponding average stresses are zero	

\begin{align}
		<\sigma_3> &=0 \\
		<\sigma_4> &=0 \\
		<\sigma_5> &=0
\end{align}
\end{subequations}
	
\vspace{2em}	
\textbf{3.2  Application to a tetragonal polydomain thin film}
\vspace{1em}

Here the case of the tetragonal domain structure is discussed in detail. The analysis of other domain structures (such as the monodomain $c$-phase and the polydomain $a/b$ and $r$ phases) is presented in the Supplemental Material. First the zero field case is considered. (At finite fields the in-plane oriented polarization vectors may tilt slightly out-of-plane, which complicates the analytical study significantly.) Since we assume that the contribution of the domain walls is constant one needs only to consider the total energy of all the domains. The total energy of the domains in a film in zero field with $c$, $b$ and $a$ domains, which are arranged in $c/a$, $c/b$ and $a/b$ subdomain structures, is then
\begin{align}
	\label{eqFcba}
	<F> - F_{DW} = <F>_{cba}=\phi_{ca}<F>_{ca} +\phi_{cb}<F>_{cb} + \phi_{ab}<F>_{ab} 
\end{align}
	
$<F>_{ca}$ is the energy of a $c/a$ domain structure with fraction $\phi_{ca}$ of the film volume and the other parameters are defined analogously. After cycling the film to a large field (`poling' the film) one expects that the $a/b$ substructure is removed and the film predominantly shows equal fractions of $c/a$ and $c/b$ domains, $\phi_{ca}=\phi_{cb}=1/2$, because of symmetry. For the same reason $F_a=F_b$ and therefore also $<F>_{ca}=<F>_{cb}=\phi F_c+(1-\phi)F_a$ \cite{remarkpoling}. Thus  \eqref{eqFcba} can be written as
 
\begin{align}
	\label{eqFcba2}
	<F>_{cba} = \phi F_c +(1-\phi)F_a
\end{align}
	
The latter result is the same as in the one-dimensional approach. Note that the final result of eq. \eqref{eqFcba2} does not depend on the specific domain structure, but only on the relative fractions $\phi$ and $(1-\phi)$ of respectively the $c$ and $a$ plus $b$ domains. The difference with the one-dimensional approach is in the macroscopic boundary conditions \eqref{eqS126BC}a,b,c, where the summation runs over the three possible domains and not only over two. 
	
The 'strict microscopic mechanical boundary conditions' on the domain walls \cite{remarkSm0}, in combination with the macroscopic boundary conditions, impose very strict limitations on the stresses, namely $\sigma_{c3}=\sigma_{a3}=\sigma_{c4}=\sigma_{a4}=\sigma _{c5}=\sigma _{a5}=\sigma_{c6}=\sigma_{a6}=0$. In the one-dimensional approach further analysis shows that $(\sigma_{c1}=\sigma_{a1}) \neq (\sigma_{c2}=\sigma_{a2})$. Applying the strict conditions in the two-dimensional approach it is found that the stress is equal in both in-plane directions $\sigma_{c1}=\sigma_{a1}=\sigma_{b1}=\sigma_{c2}=\sigma_{a2}=\sigma _{b2} \equiv \sigma $ , as one would also expect from symmetry considerations. From the `strict' electrical boundary conditions (i.e. a 90\textdegree domain wall) it follows that $P_{c3} = P_{a1} = P$. We will refer to this equality as the `\textit{strict polarization condition}'. Also under relaxed electrical boundary conditions this is a good approximation for small applied fields, which can be used to find analytical approximations for field derivatives of several parameters at $E=0$. We will see that only for zero field $P_{c3}=P_{a1}=P$ is an exact solution under relaxed electrical boundary conditions.
As discussed above we will \textit{not} use the strict mechanical domain wall boundary conditions, but only the macroscopic boundary conditions. In that case \eqref{eqS126BC} a,b can be written as

\begin{align}
	\label{Sm0}
	S^0_m=<S_1>=<S_2>= \phi_{ca}{<S_2>}_{ca}+\phi_{cb}{<S_2>}_{cb} = \frac{1}{2}\left({<S_2>}_{ca}+{<S_1>}_{ca}\right)
\end{align}

Here $<S_i>_{ca}=S_{ci}+(1-\phi)S_{ai}$ is the average strain in the $c/a$ domain structure in the in-plane directions $i=1,2$. Further use was made of the relations $<S_2>_{cb}=< S_1>_{ca}$ and $<S_1>_{cb} = <S_2>_{ca}$ that follow from symmetry considerations. From \eqref{eqS126BC}a,b,c it follows that $\phi S_{c6} + (1-\phi)S_{a6} = 0 $ and in combination with the equation of state for $S_{x6}$ it follows that ${\phi \sigma_{c6} + (1-\phi)\sigma_{a6} = 0}$. From \eqref{eqS126BC}d,e,f one has ${\phi \sigma_{ci} + (1-\phi)\sigma_{ai}= 0}$ for $i=3,4,5$. Substituting the expressions for the strain, $S_i=-{\partial G}/{\partial {\sigma }_i}$, where $G$ is given by \eqref{eqG}, into \eqref{Sm0} one arrives at the following expression for the $c$-domain fraction in the $c/b/a$ domain structure

\begin{align}
\label{eqphi} 
	\phi =\frac{[Q_{11}+Q_{12}]P^2 - 2S^0_m +(s_{11}+s_{12})(\sigma_{a1}+\sigma_ {a2})}{[Q_{11}-Q_{12}]P^2 -(s_{11}+s_{12})(\sigma_{c1}+\sigma_{c2}-\sigma_{a1}-\sigma _{a2})} 
\end{align}

Note that this result is obtained under the condition $P_{c3}=P_{a1}=P$ and all other components equal zero, valid for small (zero) fields. To make the connection with experimentally determined lattice parameters $a_{ix}=a_0(1+S_{xi})$ one can rewrite \eqref{eqphi} as

\begin{align}
\label{eqphias}
	\phi = \frac{a_{a1}+a_{a2}-2a^*_s}{a_{a1}+a_{a2}-a_{c1}-a_{c2}}
\end{align}

Thus, irrespective of the stress in the domains the domain fraction can be obtained from the \textit{measured} lattice parameters in the $c$ and $a$ (or $b$) domains. In the stress free state (for which all ${\sigma }_{xi}=0$, which is the case at zero field, as will be shown) the lattice parameters are given by $a_{c1}=a_{c2}=a_{a2}=a_{a3}=a_T$ and $a_{c3}=a_{a1}=c_T$ with $a_T$ and $c_T$ the bulk lattice parameters, so that \eqref{eqphias} reduces at zero field (sub-index 0) to

\begin{align}
\label{eqphiaT}
	\phi_0=\frac{c_T+a_T-2a^*_s}{c_T-a_T} 
\end{align}

Note that \eqref{eqphi} is valid for arbitrary polarization and stress in the domains (but without other polarization components than $P_{c3}$ and $P_{a1}$). \\
Symmetry demands that $<\sigma_1> =<\sigma_2>$. When there is no effect of the domain walls on the stress field in the domains (thus if only the macroscopic boundary conditions are used) the stress in the film must be homogeneous in both in-plane directions, $\sigma_{c1}=\sigma_{a1}=\sigma_{c2}=\sigma_{a2}=\sigma$. (One could consider the film as a strained membrane composed of connected smaller $'c'$, $'a'$ and $'b'$-oriented membranes, strained at the outer edge by the substrate.) The homogeneous in-plane stress condition also follows from numerical minimization of the energy (paragraph 4). In realistic thin films the homogeneous stress condition not necessarily applies, since local stress fields may arise from the dense domain wall structure or from defects, giving rise to inhomogeneous stress fields in the film. This is not discussed further here. With homogeneous stress \eqref{eqphi} becomes

\begin{align}
\label{eqphiQ}
 	\phi=\frac{[Q_{11}+Q_{12}]P^2-2S^0_m+2(s_{11}+s_{12})\sigma}{[Q_{11}-Q_{12}]P^2} 
\end{align}
This result is valid under the `strict polarization condition' for small but finite applied field. 

Now we write out the total energy \eqref{eqFcba2} in its components (under the strict polarization condition)

\begin{equation}
\label{Fcbax}
\begin{aligned}
<F>_{cba} &=
\sum_{x=c,a}{\phi_x \left( \alpha_1P^2_x+\alpha_{11}P^4_x+\alpha _{111}P^6_x \right)}-\phi_c EP_{c3} \\ 
	&+ 	\phi_x \left( \frac{s_{11}}{2}(\sigma^2_{x1}+\sigma^2_{x2}+\sigma^2_{x3})+
	s_{12}(\sigma_{x1}\sigma_{x2}+\sigma_{x1}\sigma_{x3}+\sigma_{x3}\sigma_{x2})+
	\frac{s_{44}}{2}(\sigma^2_{x4}+\sigma ^2_{x5}+\sigma^2_{x6}) \right) 
\end{aligned}
\end{equation}

Here $P_x$ is in the 1 and 3 direction for $x=a$ and $x=c$ respectively.
 Minimization with respect to the independent parameters $\sigma_{x4}$, $\sigma_{x5}$, and $\sigma_{x6}$ gives as simplest solutions $\sigma_{c4}=\sigma_{c5}=\sigma_{c6}=\sigma_{a4}=\sigma_{a5}=\sigma_{a6}=0$ for each domain.
(In principle constant strain solutions obeying the relations $\phi \sigma_{ci}+(1-\phi)\sigma_{ai}=0$ for $i=4,5,6$ are also allowed, giving rise to constant energy contributions $\phi s_{44}\sigma^{2}_{ci}/2$ in \eqref{Fcbax} which only cause a shift of the energy zero point, but have no effect on the electrical field dependence of the polarization and stress. We have assumed no effect of the domain walls on the strain fields in the domains, therefore all shear stresses are assumed to be zero.) Under homogeneous in-plane stress conditions and using \eqref{eqS126BC}d the terms depending on $\sigma_{x3}$ amount to $\phi s_{11}\sigma^{2}_{c3}/2$. The energy cross terms of the $c$ and $a$ domains proportional to $\sigma_{x1}\sigma_{x3}+\sigma_{x3}\sigma_{x2}$ cancel each other. Minimization of the energy with respect to $\sigma_{c3}$ makes this stress component zero. (But again \eqref{eqS126BC} allows the possibility that the domain wall introduces a constant stress (and additional strain) in the 3-direction in the adjacent domains.) \\
 The strict polarization condition imposes $P_{c3}=P_{a1}=P$ and that all other polarization components are equal zero (which is exact for $E=0$). Hence, in that case there is no polarization rotation and the $a$ and $b$ domains do not develop a 3-component under an applied field. Further on it is shown that the latter simplification leads to significant errors in the properties that depend on derivatives of the polarization and the domain fraction with respect to the applied field. With these approximations for the polarization \eqref{Fcbax} becomes 

\begin{align}
\label{Flin1}
			<F>_{cba} \approx \alpha_{1}P^2+\alpha_{11}P^4+\alpha_{111}P^6-
			\phi EP+(s_{11}+s_{12})\sigma^2 
\end{align}

Substituting \eqref{eqphiQ} in \eqref{Flin1} results into

\begin{align}
\label{Flin2}
	<F>_{cba}\approx \alpha_{1}P^2+\alpha_{11}^4+\alpha_{111}P^6-QEP
	+\frac{SE}{P}-\frac{T_1\sigma E}{P}+T_2\sigma^2
\end{align} 

where we have defined  $Q=(Q_{11}+Q_{12})/(Q_{11}-Q_{12})$, $S=2S_m/(Q_{11}-Q_{12})$, $T_1=2(s_{11}+s_{12})/(Q_{11}-Q_{12})$ and $T_2=(s_{11}+s_{12})$. 
For zero field $P_{c3}=P_{a1} \equiv P$ and the exact zero-field solutions are found from minimization of \eqref{Flin2} as

\begin{subequations}
	\label{approxsol}
	\begin{align}
	P^2(0) =P^2_s &=-\frac{\alpha_{11}}{3\alpha_{111}}
	+\left( \left( \frac{\alpha_ {11}}{3\alpha_{111}} \right)^2-\frac{\alpha_1}{3\alpha_{111}} \right)^{1/2} \\ 
	\sigma(0)&=0 \\
	\phi_0 &=\frac{[Q_{11}+Q_{12}]P^2_s-2S^0_m}{[Q_{11}-Q_{12}]P^2_s}
\end{align}
\end{subequations}

The saturation polarization value $P$ corresponds to the stress-free bulk value \cite{Haun-1} and thus eq.\eqref{approxsol}c is equivalent to that in \cite{Koukhar2001}. The result \eqref{approxsol}c is different from the result obtained for the one-dimensional approach \cite{remark1Dcase}, which is a first consequence of considering the two-dimensional domain structure. Secondly, in the two-dimensional approach the field-free domain structure resolves all stress in the film in \textit{both} in-plane directions. This is another important difference with the one-dimensional approach in which only in the direction in which a $c/a$ domain structure has developed the stress is zero, whereas parallel to the domain walls the stress is finite. We think that one may expect from symmetry considerations that the domain formation in the film gives equal responses in both in-plane directions and the one-dimensional result on the stress is therefore counterintuitive. Zero stress also implies that for $E=0$ the lattice parameters in all domains are equal to the bulk lattice parameters (provided the domain walls or other causes do not introduce additional constant stresses in the adjacent domains) and the domain fraction is given by \eqref{eqphiaT} or \eqref{eqphiQ} with zero in plane stress. 

The measured remanent polarization of the film (index \textit{f}) is

\begin{align}
	P^f_3(0)=\phi_0 P_s
\end{align} 

where $\phi(0) \equiv \phi_0$ is determined from \eqref{eqphiQ}d with $\sigma =0$ and $P=P_s$ thus from \eqref{approxsol}c. Hence from the measured remanent polarization of a (001)-oriented, epitaxial film with a tetragonal domain structure one can determine the (zero-field) $c$-domain fraction, using the bulk value for the saturation polarization. \\
 Under the strict polarization condition one finds for finite fields by minimization of \eqref{Flin2}

\begin{subequations}
	\label{solutionsapproxeq}
	\begin{align}
	-QEP^2+2\alpha_1P^3+4\alpha_{11}&P^5+6\alpha_{111}^7 =(S-T_1 \sigma)E \\  
	\sigma =\frac{T_1}{2PT_2}E &=\frac{1}{(Q_{11}-Q_{12})P}E
	\end{align}
\end{subequations}

This set of equations determines the $P-E$ and $\sigma -E$ dependencies from which the film properties in the $c/b/a$ phase can be calculated. Since the polarization only slightly changes with applied field (if no 180\textdegree domain switching takes place), thus $P\approx P_s$ the result \eqref{solutionsapproxeq} b implies that the in-plane stress in the film changes in good approximation linearly with the applied field.

\begin{figure}
	\centering
	\includegraphics[width=0.6\linewidth]{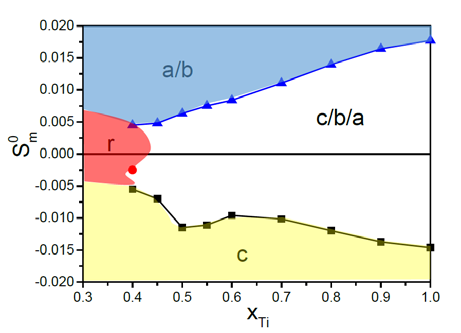}
	\caption[Phasediagram]{Phase diagram of (001)-oriented, epitaxial \ce{PbZr_{1-x}Ti_xO_3} films grown on dissimilar substrates with varying misfit strain $S_m^0$. The full data points are obtained for compositions for which material data are available. The open data points are obtained from interpolated and extrapolated material parameters. The rhombohedral $r$-phase boundary is indicative.}
	\label{Phasediagram}
\end{figure}

The boundaries of the three-dimensional tetragonal phase in zero field are obtained from \eqref{approxsol}c as $S^0_m=\frac{1}{2}(Q_{11}+Q_{12})P^2_s$ for the boundary between the $c/b/a$ and the $a/b$-phase ($\phi_0=0)$ and $S^0_m=Q_{12}P^2_s$ for the boundary between the $c/b/a$ and the $c$-phase ($\phi_0=1)$. From the numerical analysis discussed below no other phases are found for all tetragonal compositions with \ce{x_{Ti}}$\ge 0.5$. For the rhombohedral composition \ce{x_{Ti}}$=0.4$ it is found that there is only a small difference in free energy between the rhombohedral state and the $c/b/a$-phase, giving rise to switching between these phases as function of the substrate strain.  In Fig. \ref{Phasediagram} the phase diagram at room temperature as function of composition is given. Apart from the data points for \ce{x_{Ti}}$=0.4$ the rhombohedral region is indicative and obtained from extrapolation of material parameters to lower Ti content. The main differences with the earlier $c/a$ polydomain model are found around the MPB. Whereas previously additional $ca^*/aa^*$ and $ca^*/cb^*$ (in \cite{Kukhar2006} named $ca_1/ca_2$-phase) phases were found next to the tetragonal polydomain phase for the \ce{x_{Ti}}$=0.5$ and \ce{x_{Ti}}$=0.6$ compositions, these are not found for the three-dimensional, tetragonal domain structure discussed here. This is because the latter allows full stress relaxation in all in-plane directions and thus lowering of the free energy, in contrast with the two-dimensional polydomain phase which exhibits stress buildup in the direction orthogonal to the domain structure. In that case the additional phases are energetically more favorable than the \textit{c/a} phase. For the rhombohedral $x_{Ti}=0.4$ compositions qualitatively the same phase-strain dependence at room temperature is found as in \cite{Kukhar2006}.

\vspace{2em}
\textbf{3.2.1 Polarization rotation in tetragonal domains}
\vspace{1em}

For finite applied field strengths polarization rotation in the $a$ and $b$-domains occurs. To obtain accurate results for dielectric and piezoelectric properties of the film one needs to take polarization rotation into account. Following the same reasoning as for the case $E=0$ one obtains for the domain fraction

\begin{align}
\label{phiE}
	\phi(E)=
	\frac{[(Q_{11}+Q_{12})P^2_{a1}+2Q_{12}P^2_{a3}]-2S^0_m+2(s_{11}+s_{12})\sigma}
	{2Q_{12}(P^2_{a3}-P^2_{c3})+(Q_{11}+Q_{12})P^2_{a1}} 
\end{align}

which reduces to \eqref{eqphiQ} when $P_{c3}=P_{a1}=P$, $P_{a3}=0$ for $E=0$. The domain fraction $\phi(E)$ is implicitely field dependent through the field dependencies of the polarization components and the stress. We can rearrange \eqref{phiE} as

\begin{align}
\label{sigmaexact}
	\sigma =
	\frac{S^0_m-(1-\phi)\frac{Q_{11}+Q_{12}}{2}P^2_{a1}-(1-\phi)Q_{12}P^2_{a3}-\phi Q_{12}P^2_{c3}}{s_{11}+s_{12}} 
\end{align}

Now eq.\eqref{Fcbax} is rewritten in terms of the polarization components $P_{c3}$, $P_{a3}$ and $P_{a1}$ as

\begin{equation}
\begin{aligned}
\label{Ftetrag}	
	<F>_{cba} &=\phi [\alpha_1 P^2_{c3}+\alpha_{11}P^4_{c3}+\alpha_{111}P^6_{c3}-EP_{c3}]
	+(1-\phi)[\alpha_1(P^2_{a1}+P^2_{a3})+\alpha_{11}(P^4_{a1}+P^4_{a3}) \\
	&+\alpha_{11}(P^6_{a1}+P^6_{a3})+\alpha_{12}P^2_{a1}P^2_{a3}+ \alpha_{112}(P^4_{a1}P^2_{a3}+P^4_{a3}P^2_{a1})-EP_{a3}]+(s_{11}+s_{12})\sigma^2
\end{aligned}
\end{equation}

Substitution of \eqref{sigmaexact} into \eqref{Ftetrag} and minimizing with respect to $P_{a3}$ results into

\begin{align}
\label{Pa3}
	P_{a3}=\frac{E}{2(\alpha_1+\alpha_{12}P^2_s+\alpha_{112}P^4_s)}
\end{align}

Here we made use of the relations $P_{a1}(0)=P_{c3}(0)=P_s$ and the zero-field domain fraction $\phi_0$. Thus we have found an analytical expression for the out-of-plane polarization component of the in-plane oriented domains. $P_{a3}$ increases linearly with the field, or stated differently, the polarization rotates out of the plane. The analytical expressions for $P_{c3}$ and $P_{a1}$ are very complicated and also coupled and are therefore not given here. For the general case we use the numerical analysis to determine the field dependence of the different polarization components.

\vspace{2em}
\textbf{3.3  Dielectric properties of a tetragonal (001) polydomain film}
\vspace{1em}

The measured dielectric constant at zero field for a film with top and bottom electrodes under strict polarization conditions is given by

\begin{align}
\label{epstetrag}
\varepsilon_0 \varepsilon^f_{33}=\left(\pdv{P^f_3}{E}\right)_{E=0}=\left(\pdv{P}{E}\right)_0=\left( \pdv{\phi }{E}\right)_0 P_s +\phi_0 \left(\pdv{P}{E} \right)_0 
\end{align}

Here we assumed that the out-of-plane polarization is only due to the $c\ $domains, thus no polarization rotation of the $a$ (and $b$) domains. Allowing for polarization rotation under the influence of an applied field the out-of-plane polarization is given by $P^f_3=\phi P_{c3}+(1-\phi)P_{a3}$, hence

\begin{equation}
\begin{aligned}
\label{eps33rot}
	\varepsilon_0\varepsilon^f_{33rot} &=\left(\pdv{\phi}{E}\right)_0 P_s +
	\phi_0 \left( \pdv{P_{c3}}{E} \right)_0 +(1-\phi_0) \left( \pdv{P_{a3}}{E} \right)_0 \\
	&\approx a_{\phi E}P_s+\varepsilon_0\varepsilon_{a33}+\varepsilon_0(\varepsilon _{c33}-\varepsilon_{a33})([\phi_{0,0}+a_{\phi S_m}S^0_m) 
\end{aligned}
\end{equation}

Here we defined the relative permittivity of domain $x$ as $ ( \partial{P_{xi}} /\partial{E} )_0/\varepsilon_0\equiv \varepsilon_{xi3}$. In the second step of \eqref{eps33rot} we have introduced a linearization of \eqref{phiE}, 

\begin{align}
\label{linSm0}
\phi (S^0_m,E)=\phi_{0,0}+a_{\phi S^0_m}S^0_m+a_{\phi E}E
\end{align}

with $\phi_{0,0}$ the domain fraction at zero field and zero misfit strain, $ a_{\phi S^0_m}=( \partial{\phi} / \partial{S^0_m} )_{E=0} $ and  $ a_{\phi E}=( \partial{\phi} / \partial{E} )_{S^0_m=0} $. The second and third right-hand terms in \eqref{eps33rot}  amount to the domain fraction weighted average of the dielectric constants of each domain, while the first right-hand term is due to domain-wall motion. The third right-hand term in \eqref{eps33rot} is not present in \eqref{epstetrag}, since it arises from the polarization rotation in the $a$-domains. It will be seen that this term gives a significant contribution to the overall permittivity. 

In the following the various parameters in \eqref{linSm0} are determined. For zero field (hence zero stress) we have immediately from \eqref{eqphiQ}
 $ \phi_{0,0} = (Q_{11}+Q_{12})/(Q_{11}-Q_{12}) $ , earlier also defined as $Q$, and $a_{\phi S^0_m}=-2/(Q_{11}-Q_{12})P^2_s$. The values of the parameters $a_{\phi S^0_m} $ and $\phi_{0,0}$ for PZT are of the order $a_{\phi S_m}\approx -$55 and $\phi_{0,0}\approx 0.35-0.45$ for compositions in the range  \ce{x_{Ti}}$=0.5-0.6$. (For other compositions values are given in Appendix B.) Combining \eqref{solutionsapproxeq}b and \eqref{eqphiQ} a relation between the electrical field and the domain fraction is obtained, from which follows
 	
\begin{align}
\label{phiE0}
\left(\pdv{\phi}{E} \right)_0=\frac{4S^0_m\left(\pdv{P}{E}\right)_0}{(Q_{11}-Q_{12})P^3_s} +\frac{2(s_{11}+s_{12})}{(Q_{11}-Q_{12})^2P^3_s} 
\end{align}

Hence, in the approximation with no polarization rotation $a_{\phi}=( \partial{\phi} / \partial{S^0_m} )_{E=0}=2(s_{11}+s_{12})/(Q_{11}-Q_{12})^2P^3_s$. This parameter is of the order of 0.05/(100kV/cm) (100kV/cm corresponds to the electrical field range used in 1 $\mu m$ thin film applications), for tetragonal compositions in the range  $x_{Ti}=0.5-0.7$ and decreasing rapidly for higher Ti-content (see Appendix B). Thus when cycling through the polarization hysteresis loop the $c$-domain fraction changes in the order of a few percent as function of the applied field. 

\begin{figure}
	\centering
	\includegraphics[width=0.6\textwidth]{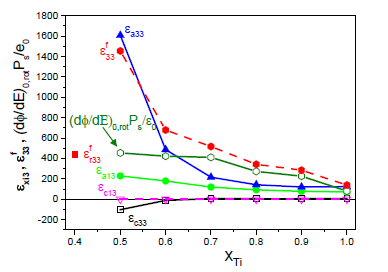}
	\caption[figeps]{Zero field relative permittivities $\varepsilon_{xi3}$ of the $c$ and $a$ domains in the $c/a/b$ phase as function of Ti-content in PZT. $\varepsilon_{33}^f$ is the net relative permittivity of the tetragonal film, while $(\partial{\phi}/\partial{E})_{0,rot}P_s/\varepsilon_0$ (eq.\ref{aphiErot}) is the contribution of domain wall motion for $S_m^0=0$. The net permittivity of PZT(\ce{x_{Ti}}$=0.6$) in the $r$-phase is given by the datapoint $\varepsilon_{r33}^f$. All results are obtained taking polarization rotation in the domains under influence of a field in the 3-direction into account.}
	\label{figeps}
\end{figure}

If one takes polarization rotation into account in Eq. \eqref{eps33rot} one needs values for the permittivities $\varepsilon_{xi3}$. The permittivity $\varepsilon_{a33}$ of the $a$-domains is simply obtained from \eqref{Pa3} as

\begin{align}
	\varepsilon_0 \varepsilon_{a33} = \left( \pdv{P_{a3}}{E} \right)_0 =
	\frac{1}{ 2(\alpha_1+\alpha_{12}P^2_s+\alpha_{112}P^4_s) } 
\end{align}

and is thus independent of the strain $S^0_m$.
Although analytical expressions for the zero-field permittivities $\varepsilon_{a13}$ and $\varepsilon_{c33}$ can be obtained, these are coupled and analytical results become cumbersome. Therefore we use the numerically obtained values from energy minimization of eq.\eqref{Ftetrag} (discussed further in section 4). In the $c/b/a$ phase the permittivities $\varepsilon_{xi3}$ are independent of the strain $S^0_m$. This is because these permittivities are intrinsic properties and therefore do not depend on the domain fraction, which is determined by the strain. This can also be seen by writing the permittivity as $\varepsilon_0 \varepsilon_{xi3} = (\partial{P_{xi}}/\partial{E}_0 =  \sum_{j=1,2,k=1,2,3} (\partial{P_{xi}}/\partial {S_k})_0 (\partial{S_k}/\partial{\sigma_j})_0 (\partial{\sigma_j}/\partial{E})_0  $ . The first two partial derivatives only depend on the intrinsic properties of the unit cell, while the last derivative, that connects the stress in the unit cell to the field in the film, is given by \eqref{solutionsapproxeq}b, which is also independent of the film strain. Fig.\ref{figeps} shows the numerically calculated permittivities of the different domains in the $c/b/a$ phase (for $x_{Ti}\ge 0.5$) and for the rhombohedral $r$- phase. It is seen that $\varepsilon_{a33}$ increases rapidly when decreasing the Ti-content on approaching the MPB and so does the relative dielectric constant of the film. Surprisingly the $\varepsilon_{c33}$ decreases from zero to slightly negative towards the MPB, whereas the in-plane permittivity $\varepsilon_{a13}$ is positive and only slightly increases with decreasing Ti-content. 

Naively one would expect that the $\partial{P^f_3}/\partial{E}$ dependence of the dielectric constant of the film arises from the $c$-domains and this is also the result of the polarization approximations leading to \eqref{epstetrag}. However for compositions close to the MPB this dependence is nearly fully due to the out-of-plane rotation of the polarization vector of the $a$-domains. For example in the case of PZT(\ce{x_{Ti}}$=0.6$) one has $\varepsilon_{c33}\approx -12$, $\varepsilon _{c13}\approx 0$, $\varepsilon_{a33}\approx 487$ and $\varepsilon_{a13}\approx 179$. The contribution of the polarization extension in the $c$-domains and polarization rotation in the $a$-domains to the relative dielectric constant is thus (for $S^0_m=0$) $ \varepsilon ^f_{33rot}(0,0) - a_{\phi E}P_s/\varepsilon _0 = \varepsilon_{a33}+(\varepsilon_{c33}-\varepsilon_{a33})\phi_{0,0}=254 $ , which is of the order of the domain wall motion contribution $ a_{\phi E,rot}P_s/\varepsilon_0=422 $, thus $\varepsilon^f_{33rot}(\ce{x_{Ti}})=0.6=676$. We note that the large contribution of the latter implies that one can expect a significant effect of domain wall pinning in \textit{AC}-measurements. The other terms arise from intrinsic susceptibility contributions and depend on the static $P-E$ loop only. This may explain the often observed discrepancy between the dielectric constant versus applied field  loops calculated from the quasi $DC$-measurement of the $P-E$ loop and determined from $C-V$ measurements.\
			
Fig.\ref{figeps} also shows the net relative dielectric constant of the film, $\varepsilon ^f_{33}$, as function of the composition, showing a sharp upturn when approaching the MPB, where the out-of-plane rotation of the $a$-domain polarization translates into a large permittivity contribution. The positive value of $\varepsilon_{a13}$ implies that also the in-plane polarization of the $a$-domain increases, which is due to the in-plane stress increasing with the applied field. Maybe somewhat surprisingly $\varepsilon_{c33}$ is very small and even slightly negative close to the MPB, implying that the $c$-domain polarization decreases with increasing field. This is again due to the increasing in-plane stress and strain, which translates through the Poisson effect into a shortening of the long axis. Since we have assumed that there is no in-plane polarization component in the $c$-domain (no domain tilt), there is no out-of-plane rotation component and $\varepsilon_{c13}$ is therefore zero.

The lines in Fig.\ref{figeps} are a guide to the eye. $\varepsilon_{r33}$ and $\varepsilon_{r13}$ are the relative permittivities of the rhombohedral phase of PZT(60/40) for  $E=S^0_m=0$.
The above approximation \eqref{phiE0} for the change of the domain fraction with applied field is not accurate enough when polarization rotation is important. Taking polarization rotation into account in the expression for the domain fraction \eqref{phiE} one finds

\begin{align}
\label{dphidErot}
	\left( \pdv{\phi}{E} \right)_{0,rot} = (1-\phi_0) \frac{2(Q_{11}+Q_{12})} {(Q_{11}-Q_{12})P_s}
	\varepsilon_0\varepsilon_{a13}+
	\phi_0\frac{4Q_{12}} {(Q_{11}-Q_{12})P_s}\varepsilon_0\varepsilon_{c33}
	+\frac{2(s_{11}+s_{12})} {(Q_{11}-Q_{12})^2P^3_s} 
\end{align}

The linearization coefficient $a_{\phi E}$ is then obtained by substituting $\phi_{0,0}$ for $\phi_0$ as

\begin{align}
\label{aphiErot}
a_{\phi E,rot} \approx 
\frac{-4Q_{12}(Q_{11}+Q_{12})} {(Q_{11}-Q_{12})^2P_s} \varepsilon_0 (\varepsilon_{a13}-\varepsilon_{c33})+ 
\frac{2(s_{11}+s_{12})} {(Q_{11}-Q_{12})^2P^3_s}
\end{align}
 
In Fig.\ref{figeps} the contribution to the relative dielectric constant arising from domain wall motion, $a_{\phi E,rot}P_s/\varepsilon_0$, is plotted as function of the composition. It is seen that the contribution of domain wall motion to $\varepsilon^f_{33rot}$ is dominant for \ce{x_{Ti}}$>0.65$, while for lower Ti-content the polarization rotation of the $a$-domains gives the largest contribution to the film permittivity. 

\begin{figure}
	\centering
	\includegraphics[width=1.05\linewidth]{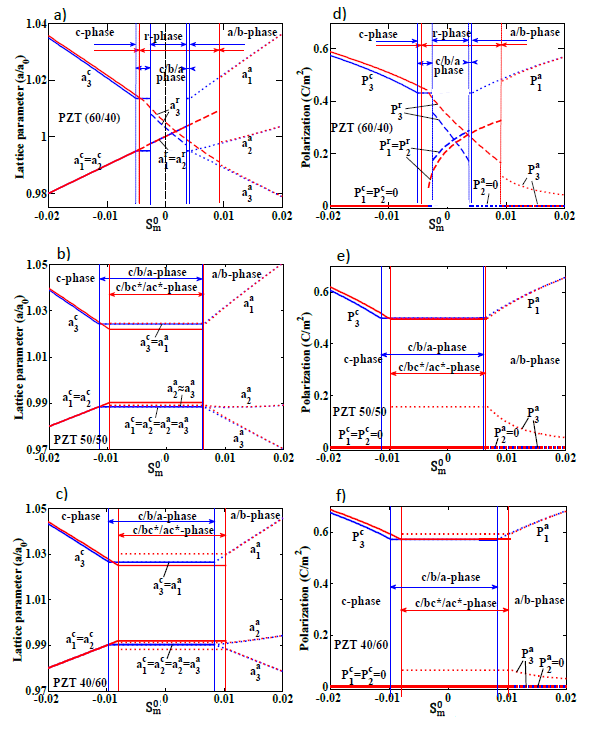}
	\caption[Lattice parameters]{Lattice parameters (a-c) and polarization components (d-f) as function of misfit strain $S_m^0$ at applied fields $E=0$ (blue) and $E=200$ kV/cm (red), for PZT(\ce{x_{Ti}}$=0.4$) (a,b)  PZT(\ce{x_{Ti}}$=0.5$) (c,d) and PZT(\ce{x_{Ti}}$=0.6$) (e,f)}
	\label{lattice-polarization}
\end{figure}

\vspace{2em}
\textbf{3.4  Lattice parameters of a tetragonal (001) polydomain film}
\vspace{1em}

The lattice strains as function of electric field in the $c/b/a$ phase are obtained as (using relaxed polarization conditions)

\begin{subequations}
	\label{strains}
	\begin{align}
	S_{c1} &=S_{c2} =(s_{11}+s_{12})\sigma +Q_{12}P^2_{c3} &\approx \frac{(s_{11}+s_{12})}{(Q_{11}-Q_{12})P_s}E+Q_{12}P^2_s+{2Q}_{12}\varepsilon_{c33}P_sE \\
	S_{a1} &=(s_{11}+s_{12})\sigma +Q_{11}P^2_{a1}+Q_{12}P^2_{a3} &\approx \frac{(s_{11}+s_{12})}{(Q_{11}-Q_{12})P_s}E+Q_{11}P^2_s+2Q_{11}\varepsilon_{a13}P_sE \\
	S_{a2} &=(s_{11}+s_{12})\sigma +Q_{12}P^2_{a1}+Q_{12}P^2_{a3} &\approx \frac{(s_{11}+s_{12})}{(Q_{11}-Q_{12})P_s}E+Q_{12}P^2_s+2Q_{12}\varepsilon _{a13}P_sE \\
	S_{c3} &=2s_{12}\sigma +Q_{11}P^2_{c3} &\approx \frac{2s_{12}}{(Q_{11}-Q_{12})P_s}E+Q_{11}P^2_s+2Q_{11}\varepsilon_{c33}P_sE \\
	S_{a3} &=2s_{12}\sigma +Q_{12}P^2_{a1}+Q_{11}P^2_{a3} &\approx \frac{2s_{12}}{(Q_{11}-Q_{12})P_s}E+Q_{12}P^2_s+2Q_{12}\varepsilon_{a13}P_sE \\
	S_{c4}&=S_{c5}=S_{c6}=S_{a4}=S_{a5}=S_{a6}=0
	\end{align}
\end{subequations}
\vspace{1em}

The lattice parameters follow from the relations $a_{xi}=a_0(1+S_{xi})$. Note that the strains only change with the applied field and are independent of the substrate strain $S^0_m$, since the relative permittivities are not changing with $S^0_m$ in the $c/b/a$ phase. Because $\varepsilon_{c33}$ is very small the short axes of the $c$-domain unit cells, $a_{c1}=a_{c2}$, increase with increasing field due to the term $(s_{11}+s_{12})E/(Q_{11}-Q_{12})P_s$, caused by the stress. The long axis $a_{c3}$ decreases in length due to both the negative stress term, $2s_{12}E/(Q_{11}-Q_{12})P_s$ and the negative piezoelectric effect $2Q_{11}\varepsilon_{c33}P_sE$ because $\varepsilon_{c33}<0$. For all $c$-domain lattice parameters the length changes are dominated by elastic effects caused by the in-plane field induced stress and not by the intrinsic piezoelectric effect. The long, in-plane axis $a_{a1}$ of the $a$-domain is elongated by the stress and the piezoelectric effect, $2Q_{11}\varepsilon _{a13}P_sE$ since $\varepsilon_{a13}$ has a significant positive value. The two $a$-domain short axes respond differently to the stress, but both shorten by the piezoelectric effect, $2Q_{12}\varepsilon_{a13}P_sE$ again because $\varepsilon_{a13}>0$. In Fig.\ref{lattice-polarization}a-c the lattice parameters are shown as function of the substrate strain for the different phases for PZT(\ce{x_{Ti}}$=0.4$), PZT(\ce{x_{Ti}}$=0.5$) and PZT(\ce{x_{Ti}}$=0.6$) for zero field and $E$=200 kV/cm.

\begin{figure}
	\centering
	\includegraphics[width=0.95\linewidth]{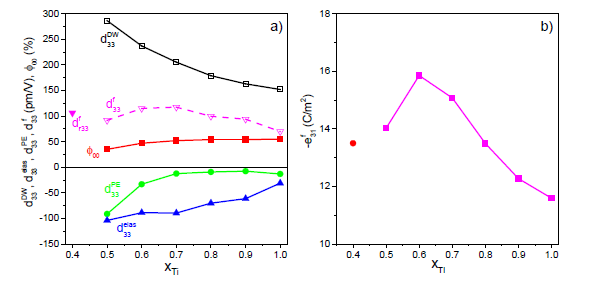}
	\caption[Polarization rotation]{a) Contributions to the net piezoelectric constant $d_{33}^f$ of a clamped film in the $c/a/b$ phase as function of the composition: domain wall motion contribution $d_{33}^{DW}$, elastic contribution $d_{33}^{elas}$ and piezoelectric contribution $d_{33}^{PE}$ , for $E=0$ and $S_m^0=0$. The net piezoelectric coefficient of PZT(\ce{x_{Ti}}$=0.6$) in the $r$-phase is given by the datapoint $d_{r33}^f$. b) Piezoelectric coefficient $e_{31}^f$ of a clamped film in the $c/b/a$ phase. The red data point gives the value of the rhombohedral PZT(\ce{x_{Ti}}$=0.4$) composition for $S_m^0$ and $E=0$ kV/cm.}
	\label{figd33e31}
\end{figure}

\vspace{2em}
\textbf{3.5  Piezoelectric properties of a tetragonal, polydomain (001)-film}
\vspace{1em}

The average out-of-plane strain is $<S_3> =\phi S_{c3}+(1-\phi)S_{a3}$, so that the effective piezoelectric parameter $d^f_{33}$ at zero field and constant stress, under the strict polarization conditions, is given by

\begin{align}
\label{d33f}
d^f_{33}=
\left( \pdv {<S_3>}{E} \right) ^\sigma _0 = \left( \pdv {\phi}{E} \right)_0 (Q_{11}-Q_{12})P^2_s+2\phi_0 Q_{11}P_s\left(\pdv{P}{E}\right)_0 +\frac {2s_{12}}{(Q_{11}-Q_{12})P_s}
\end{align}

Allowing for polarization rotation an extra term arises due to polarization rotation and the associated piezoelectric effect in the $a$-domain

\begin{align}
\label{d33frot}
d^f_{33rot}\approx \left(\pdv{\phi}{E}\right)_{0,rot}(Q_{11}-Q_{12})P^2_s+
\varepsilon_0 \left(\phi_0Q_{11}\varepsilon_{c33}+(1-\phi_0)Q_{12}\varepsilon_{a13}\right){2P}_s
+\frac{2s_{12}}{(Q_{11}-Q_{12})P_s}
\end{align}

With $d_{c33}\equiv 2Q_{11}\varepsilon_0\varepsilon_{c33}P_s$, $d_{a31}\equiv 2Q_{12}\varepsilon_0\varepsilon_{a13}P_s$ the relevant piezoelectric coefficients of the $c$ and $a$-domains, this can for $S^0_m=0$ also be written as

\begin{align}
\label{d33frot2}
d^f_{33rot,0} \approx a_{\phi E,rot} (c_T-a_T)+\left( \phi_0d_{c33}+(1-\phi_0)d_{a31} \right)+ \frac{2s_{12}P_s} {(c_T-a_T)}
\end{align}

The first right-hand term in \eqref{d33frot} and \eqref{d33frot2}, arising from domain wall motion (further on named $d^{DW}_{33}$), is a measure of the change of the domain fraction weighting of the (zero-field) out-of-plane lattice parameters with changing field. The second piezoelectric term in \eqref{d33frot2} and partly missing in \eqref{d33f} is the contribution of the piezoelectric effects of the $a$-domains to the piezoelectric term, $d^{PE}_{33}=\phi_0d_{c33}+(1-\phi_0)d_{a31}$. 
The last, elastic term, $d^{elas}_{33}=2s_{12}P_s/(c_T-a_T)$, arises from the field dependence of the lattice parameters through the changing stress and is negative: with increasing field the in-plane tensile stress increases and contracts the out-of-plane lattice parameters through the Poisson effect. The signs of both $d_{c33}$ and $d_{a31}$ and thus of $d^{PE}_{33}$, as well as of $d^{elas}_{33}$ are negative. 
Thus only the domain wall motion gives rise to a positive piezoelectric constant, but its effect is counteracted by the intrinsic piezoelectric effect, dominated by the contraction of the $a$-domain in the 3-direction, and stress buildup in the film. $d^{PE}_{33}$ and $d^{elas}_{33}$ do not depend on (frequency dependent) domain wall motion, whereas the first term does. Therefore we expect that the value given by \eqref{d33frot} (or \eqref{d33frot2}) poses an upper limit of $d^f_{33}$, while with increasing domain wall pinning $d^f_{33rot}$ can decrease significantly. In fig. \ref{figd33e31}a the various contributions to the total piezoelectric coefficient $d^f_{33rot}$ are plotted as function of the composition for $S^0_m=0$. It appears that the piezoelectric coefficient of the film does hardly vary over the tetragonal composition range, i.e. only between 87 and 115 pm/V, with a shallow maximum for $x_{Ti}\approx 0.6$. We will see furtheron that the variation becomes somewhat larger close to the $c/b/a - a/b$ phase transition at $\phi_0=0$, which takes place at a larger strain value.

All other average strains and strain derivatives with respect to the field are zero, therefore all other piezoelectric coefficients $d_{ij}\left(i,j\neq 3\right)$=0. There is no obvious relation of eq.\eqref{d33frot} with the usual expression $d^f_{33}=d_{33}-{2s_{13}d_{31}}/{\left(s_{11}+s_{12}\right)}=2\left(Q_{11}-{2s_{13}Q_{12}}/{\left(s_{11}+s_{12}\right)}\right){\varepsilon }_{33}P_s$ for a clamped film \cite{Lefki1994,Muralt1996}. In the Supplemental Material it is shown that the latter expression is valid exactly only for a homogeneous (single domain) $c$-oriented tetragonal clamped film. From the discussion above it is obvious that polarization rotation, domain fractions and domain wall motion have to be taken into account in the polydomain phase. It appears coincidental that the numerical value of the latter expression, using the values for unstrained PZT(\ce{x_{Ti}}$=0.6$), resulting in $d^f_{33}$=105 pm/V, is close to the result obtained from the full model discussed here. For other compositions the discrepancy is much larger.

The in-plane stress components at low field are given by \eqref{solutionsapproxeq}b $\sigma_{c1}=\sigma_{a1}=\sigma _{c2}=\sigma_{a2}=\sigma =E/(Q_{11}-Q_{12})P_s$. All other stress components are zero. Thus the non-zero piezoelectric coefficients $e^f_{3i}$ of the clamped film in the polydomain $c/b/a$ state are obtained as

\begin{align}
\label{e31f}
e^f_{31} = e^f_{32} = -\left( \pdv {\sigma_1}{E} \right)^S_0= \frac{-1}{(Q_{11}-Q_{12})P_s}
\end{align}

There is no dependence on the domain fraction, nor on the strain, since only intrinsic piezoelectric properties determine this piezoelectric coefficient in the polydomain phase. In fig.\ref{figd33e31}b $e^f_{31}$, calculated from \eqref{e31f},  is plotted as function of the composition. Maybe somewhat surprisingly the tetragonal compositions show a shallow maximum for PZT(40/60). Also is shown the value for the rhombohedral composition PZT(60/40) if it were forced in the tetragonal polydomain phase by substrate induced strain. This value is significantly larger than the maximum value of the tetragonal compositions. The usual expression for $e_{31}$ of a clamped thin film is $e_{31}=d_{31}/(s_{11}+s_{12})$ \cite{Lefki1994,Muralt1996}. Again the difference between these expression arises from the fact that \eqref{e31f} is the result of considering in detail the domain distribution of the film and expressing all film parameters in terms of microscopic properties.

\vspace{2em}
\noindent \textbf{4 Numerical analysis}
\vspace{1em}

A numerical analysis was performed for the different phases in which the film can organize its domains. The phase with the minimum total energy at given temperature, applied field and misfit strain is considered to be the phase in which the film is organized. For the $c/ac*/bc^*$-phase the two free energy functions for the $c$ and $a$ domain are coupled by the domain fraction as free parameter. A prioiri we have set the stresses $\sigma_{x4,5,6}$ to zero, based on the analysis in section 3. The polarization vector components and the in-plane stresses of the $c$ and $a$ domains were taken as independent free parameters. The energy minimization of \eqref{eqFcba2} gives solutions for all stress and polarization vector components and the domain fraction $\phi$. The in-plane stress components are found to be equal, as was already argued in the analysis above, and the out-of-plane stress components in each domain type, $\sigma_{x3}$, are equal zero. Note that this is a more stringent result than the boundary condition $<\sigma_3>=0$. The latter condition in principle allows for a solution with non-zero constant stresses in the different domains, which would give a constant energy contribution to the free energy. 

The polydomain phase goes over into the single domain $c$-phase for $\phi=1$, into the polydomain $ac^*/bc^*$-phase at $\phi=0$ and into the paraelectric $p$-phase, when the polarization components are equal 0. To describe the rhombohedral $r$-phase only one free energy function is needed because of symmetry (see Supplemental Material), with free parameters $P_{r1}=P_{r2}\neq 0$ and $P_{r3}$. The latter polarization component is not necessarily equal to the in-plane components and polarization rotation occurs under varying misfit strain and applied field. Again we assume $\sigma_{r4,5,6}=0$ as follows from analytical minimization, as well as the symmetry condition $\sigma_{r1}=\sigma_{r2}$. 
\begin{sloppypar}
The film properties are calculated as respectively 
$\varepsilon^f_{33}=\left( <P_3(\delta E)>-<P_3(0)> \right)/\delta E$, 
$\varepsilon_{xi3}=\left( <P_{xi}(\delta E)>-<P_{xi}(0)> \right)/\delta E$,
$d^f_{33}=\left( <S_3(\delta E)>-<S_3(0)> \right)/\delta E$,
$e^f_{31}=\left( \sigma_1(\delta E)>-\sigma_1(0)> \right) /\delta E$,
with $\delta E=1$ kV/cm
\end{sloppypar}

\vspace{2em}
\begin{figure}
	\centering
	\includegraphics[width=0.8\linewidth]{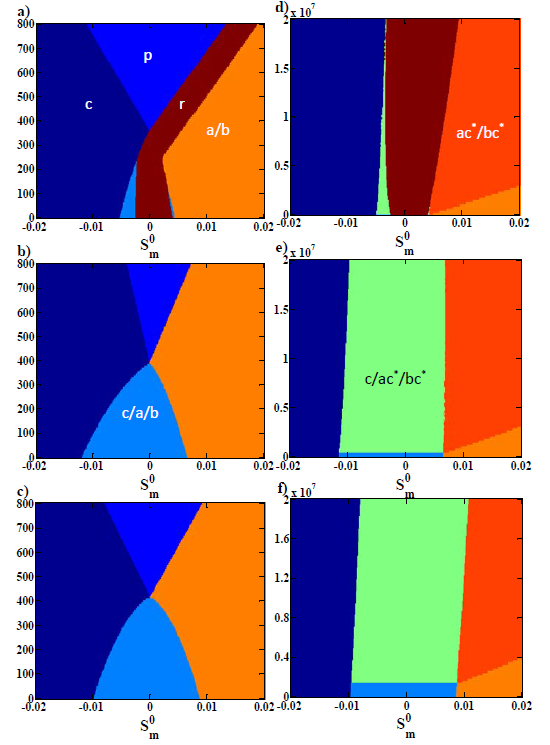}
	\caption[TS-phasediagrams]{a,d) Temperature-strain and field-strain phase diagrams of PZT(\ce{x_{Ti}}$=0.4$); b,e) idem PZT(\ce{x_{Ti}}$=0.5$); c,f) idem PZT(\ce{x_{Ti}}$=0.6$). $p$-phase is middle blue; $c$-phase is dark blue; $c/a/b$-phase is light blue; $r$-phase is brown; $a/b$-phase is orange; $ac^*/bc^*$-phase is red; $c/ac^*/bc^*$-phase is green. The strain range accessible with usual substrates (with thermal expansion coefficients in the range 0-11.5 ppm/K) is $S_m^0$= -0.0027 to 0.0039 ; -0.0037 to 0.0029 ; -0.0022 to 0.0044  for PZT(\ce{x_{Ti}}$=0.4, 0.5, 0.6$ resp.), see Appendix A.}
	\label{figtemp-strain-phasediagrams}
\end{figure}

\textbf{4.1  Numerical results}
\vspace{1em}

Here we compare the results for the rhombohedral compositions PZT(\ce{x_{Ti}}$=0.4$), the PZT(\ce{x_{Ti}}$=0.5$) composition close to the MPB and the tetragonal composition PZT(\ce{x_{Ti}}$=0.6$). Fig.\ref{figtemp-strain-phasediagrams}a-c shows the temperature-strain ($T-S^0_m)$ phase diagrams at zero field. As one would expect the $c$-phase is obtained for large compressive substrate induced stress (very negative misfit strain) and the $a/b$ phase for large tensile substrate induced stress (very positive misfit). For the intermediate misfit values the stress-free polydomain $c/b/a$ phase arises. For the rhombohedral composition this phase competes with the $r$-phase, which has nearly the same energy, resulting into the $r$-phase over a large part of the polydomain misfit strain range. Contrary to the one-dimensional approach, which resulted into several additional phases, simple phase diagrams are found for the tetragonal compositions with only $c$, $c/a/b$ and $a/b$ phases and the paraelectric phase at high temperatures. The $c-c/b/a$ phase boundaries in the $T-S^0_m$ diagram of the tetragonal as well as the rhombohedral compositions are defined by $\phi_0=1$ in \eqref{approxsol}d resulting in the relation $S^0_m=Q_{12}P^2_s$  and the $c/b/a-a/b$ phase boundary is defined by $\phi_0=0$ giving $S^0_m=\frac{1}{2}(Q_{11}+Q_{12})P^2_s$. In the case of PZT(\ce{x_{Ti}}$=0.6$) the $r$-phase largely replaces the $c/b/a$ phase. The latter phase only occurs at lower temperatures close to the phase boundaries with the $c$ and $a/b$ phases, forming a transition phase between the $r$-phase and the $c$ and $a/b$ phases. The range of substrate misfit strains $S^0_m$ accessible with substrates with thermal expansion coefficients in the range -11.5 to 0 ppm/K are $S^0_m=-0.0028...0.0040$ for PZT(\ce{x_{Ti}}$=0.6$); $\ -0.0038...0.0029\ \ $ for PZT\ce{x_{Ti}}$=0.5$) and $-0.0023...0.0044\ $ for PZT(\ce{x_{Ti}}$=0.4$), respectively (see appendix B for details). It is seen that on usual substrates the domain structure tends to be in the $c/b/a$ phase for $x\ge 0.5$ and in the $r$-phase for the rhombohedral compositions. Thus the model predicts no other phases then are present in the bulk phase. 

The field-strain ($E-S^0_m$) phase diagrams (Fig.\ref{figtemp-strain-phasediagrams}d-f) at room temperature show that the applied field poses a relatively small force on the system, since the phase boundaries between the $c$, $c/b/a$ and $a/b$ phases are nearly vertical, except for the transition between the $a/b$ and $r$ phases. The latter is an indication that under the influence of the field the polarization rapidly rotates out of plane, changing the $ac^*/bc^*$ structure into an $r$-phase. In principle the $c/b/a$ and the $a/b$ phase only exist for $E=0$. The border between the $c/b/a$ and $c/bc^*/ac^*$ respectively $a/b$ and $ac^*/bc^*$ phases are drawn here (somewhat arbitrarily) for $ac^*$ (or $bc^*$)-components that are 1\% of the in-plane components. It is seen that for the MPB and tetragonal composition with increasing tetragonality and in-plane strain a higher field is needed to drive the polarization vector of the in-plane oriented domains out of plane.

\begin{figure}
	\includegraphics[width=0.95\linewidth]{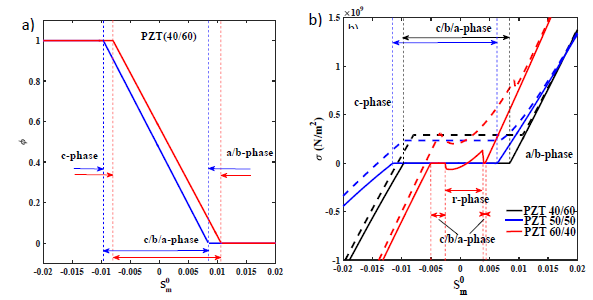}
	\caption[domain fraciton]{a) $c$-domain fraction $\phi$ for the tetragonal PZT(\ce{x_Ti}$=0.6$) composition for $E=0$ (blue) and $E=200$  kV/cm (red dashed) as function of substrate induced strain $S_m^0$. b) In-plane stress $\sigma$ for $E=0$ (solid) and $E=200$  kV/cm (dashed) as function of substrate induced strain $S_m^0$, for rhombohedral PZT(\ce{x_{Ti}}$=0.4$), near-MPB PZT(\ce{x_{Ti}}$=0.5$) and tetragonal PZT(\ce{x_{Ti}}$=0.6$)}
	\label{phi-sigma}
\end{figure}

In Fig.\ref{phi-sigma}a the $c$-domain fraction (for the PZT($x_{Ti}=0.6$) and the in-plane stress versus the misfit strain are given for zero and large field (200 kV/cm). The domain fraction changes linearly with the misfit strain, as was found analytically as well. With increasing field its value increases for the same $S^0_m$ value. Fig.\ref{phi-sigma}b shows that all in-plane stress in the zero-field $c/b/a$ phase is resolved by changing the domain fraction, as was also found analytically. The (absolute) stress rapidly increases when $\phi $ reaches its limits 0 and 1 and the domain structure becomes respectively the $a/b$ phase and $c$ phase. For finite fields the stress increases in accordance with \eqref{solutionsapproxeq}b, reducing the energy gain by the electric field terms in \eqref{Fcbax}. In the $r$-phase the stress is mostly non-zero, but the polarization rotation is used to decrease the stress, reducing the elastic energy and thus minimizing the total free energy. 

The components of the polarization in the various domains for zero and large field as well as the lattice parameters are shown in fig.\ref{lattice-polarization} as function of substrate induced misfit strain. For \ce{x_{Ti}}$=0.5$ the lattice parameters and the polarization components in the $c/b/a$-phase at zero field do not depend on the strain $S^0_m$, because the stress is zero, as predicted by \eqref{strains}. This demonstrates that in this phase the elastic energy can be nullified by shifting the domain walls. The change of the lattice parameters with applied field is due to both the piezoelectric effect and the increasing in-plane stress. It is seen that the long axis of the $c$-domain decreases with field, while the short axes increases. The change in lattice parameters can be interpreted with the relations \eqref{strains}, showing the role of the piezoelectric effect and the field-induced stress. Fig.\ref{lattice-polarization} clearly show the significant out-of-plane rotation of the $a$-domain polarization vector under an applied field. This demonstrates that the strict polarization condition (the polarization vectors in the various tetragonal domains have the same length and do not rotate under the applied field) is indeed a very strong restriction, which in realistic films with compositions close to the MPB are likely not to hold. It is also observed that the $c$-domain polarization hardly changes in magnitude under an applied field. 

\begin{figure}
	\centering
	\includegraphics[width=0.95\linewidth]{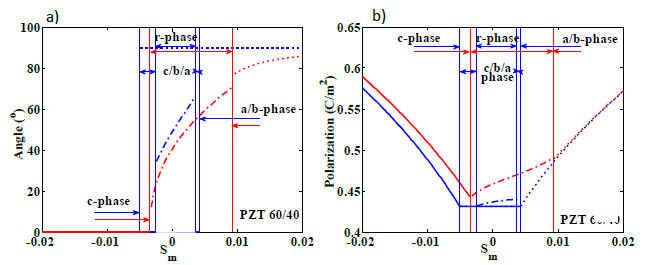}
	\caption[Polarization rotation]{a) Polarization rotation and b) polarization extension in the $r$-phase of a PZT(\ce{x_{Ti}}$=0.4$) thin film for $E=0$ (blue) and $E=200$ kV/cm (red) as function of the misfit strain $S_m^0$.}
	\label{pol rotation}
\end{figure}

\begin{sloppypar}
In the $r$-domain polarization rotation is the main mechanism to adapt to varying stress or field conditions, which can be visualized better by writing the polarization vector as $\overrightarrow{P}=P_s(sin\theta ,sin\theta ,cos\theta)$ with $\theta $ the field and strain dependent angle of the polarization vector with (field and strain dependent) length $P_s$ with the film normal. In Fig.\ref{pol rotation}a it is seen from the zero-field curve that the polarization in the $r$-phase rotates under the influence of the misfit strain towards the film plane, but that the polarization value is fairly constant (Fig.\ref{pol rotation}b). Under an applied field polarization rotation dominates the change in the polarization components causing the polarization to rotate towards the out-of-plane direction. The vector length $P_s$ increases only in the order of a few percent for large fields. Further the polarization direction jumps abruptly at the phase boundaries, whereas the polarization length varies (nearly) continuously with changing strain. In the $c$-phase there is a small unit cell extension with increasing field related to the small polarization vector length increase of the order of 2-3\%. In the $a/b$-phase the vector length only depends on the misfit strain but not on the field. For large field the boundary with the $r$-phase shifts considerably along the strain axis and the vector length becomes very sensitive to the applied field over the range of the shift, increasing with up to 8\% for $\Delta E$= 200 kV/cm. 
\end{sloppypar}

\begin{figure}
	\centering
	\includegraphics[width=1\linewidth]{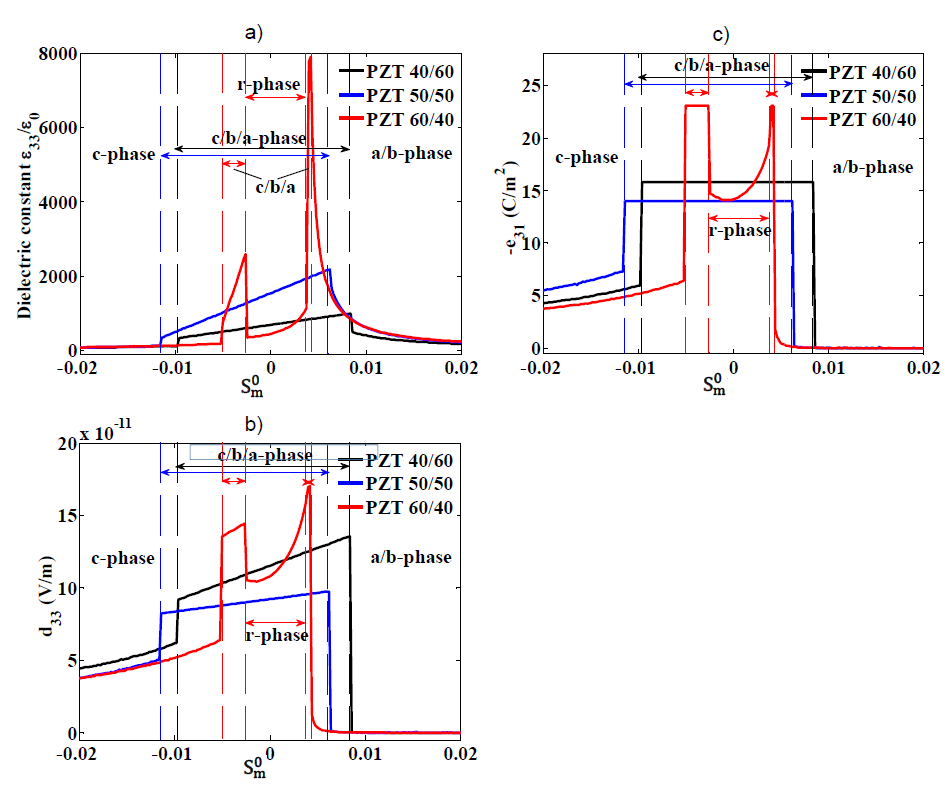}
	\caption[properties]{Room temperature relative dielectric constant $\varepsilon_{33}^f/\varepsilon_0$ (a). Piezoelectric coefficients $d_{33}^f$ (b) and $e_{31}^f$ (c) of thin films for PZT(\ce{x_{Ti}}$=0.4$) in red (a,b) PZT(\ce{x_{Ti}}$=0.5$) in blue (c,d) and PZT(\ce{x_{Ti}}$=0.6$) in black, at zero field as function of misfit strain $S_m^0$.}
	\label{properties}
\end{figure}

\vspace{1em}
The film relative dielectric constant as function of strain is shown in Fig.\ref{properties}a. In the $c/b/a$ phase of PZT(\ce{x_{Ti}}$=0.5$) and PZT(\ce{x_{Ti}}$=0.6$) the $\varepsilon^f_{33}$ increases linearly with the strain, which is due to the domain fraction dependence of the coefficients of the intrinsic permittivities of the $a$ and $c$ unit cell term in \eqref{eps33rot}. The domain wall motion gives a significant, constant contribution to the relative dielectric constant . This also explains the abrupt decrease in $\varepsilon^f_{33}$ at the $c/b/a$ phase boundaries, where the domain wall contribution suddenly drops to zero. Eq.\eqref{eps33rot} indicates that the polarization change in the $c$-domains is so small that the main intrinsic contribution to $\varepsilon^f_{33}$ arises from the $a$-domains. For the case of the $r$-domain one can write the dielectric constant alternatively in terms of polarization rotation and extension, $\varepsilon _0\varepsilon ^f_{33}=$ $(\partial{P_3}/\partial{E})_0$ $=(\partial {P_s}/\partial{E} )_0cos{\theta_0}-P_{s0}sin{\theta_0}$ $(\partial{\theta} /\partial{E})_0$ $\equiv \varepsilon_0(\varepsilon^{ext}_{33}+\varepsilon^{rot}_{33})$. Here $P_{s0}$ and $\theta _0$ are the polarization length and angle at zero field and given strain.
From numerical analysis it can be shown that for PZT(\ce{x_{Ti}}$=0.4$) the ratio of the contributions of polarization rotation and extension is $\varepsilon^{rot}_{33}/\varepsilon^{ext}_{33}\approx 3.2$ for $S^0_m=0$, thus $\varepsilon^f_{33}$ is dominated by polarization rotation.

The piezoelectric coefficient $d^f_{33}$ of the tetragonal compositions, plotted in Fig.\ref{properties}b, can most easily be interpreted in terms of \eqref{d33frot}. The misfit dependence of $d^f_{33}$ is due to the domain fraction weighted contributions of the intrinsic piezoelectric coefficients, whereas the constant part is due to domain wall motion and the (constant) induced strain. Again we see that the sharp drop of $d^f_{33}$ at the $c/b/a$ phase boundaries is due to the disappearance of the field sensitive domain fraction. 
For the $r$-domain we can write the piezoelectric constant again in terms of the polarization length and angle 
$d^f_{33}=2s_{12}(\partial{\sigma}/\partial{E})_0+2P_{s0}(\partial{P_s}/\partial{E})_0
(Q_{11}cos^2{\theta_0}+Q_{12}sin^2{\theta_0})-(\partial{\theta}\partial{E}_0P^2_{s0}
(Q_{11}-Q_{12})sin{2\theta_0}$. 
For PZT(\ce{x_{Ti}}$=0.5$) the first term is due to elastic effects ($d^{elas}_{33}=-84 $ pm/V for $S_m=E=0$), 
while the second ($d^{ext}_{33}=16 $pm/V) and third term ($d^{rot}_{33}=149 $pm/V) are due to polarization extension and rotation respectively. 
Thus the piezoelectric effect is dominated by polarization rotation, while the polarization extension gives a 
small contribution ($d^{rot}_{33}/d^{ext}_{33}=9.3 $). The elastic term causes a large counteracting effect.
The numerical results for the piezoelectric coefficient $e^f_{31}$ of the tetragonal compositions shown in Fig.\ref{properties}c are well described by \eqref{e31f}. 
\vspace{1em}

In the above model the minimum energy solution for the (near) MPB composition is a polydomain tetragonal phase. Experimentally it was found that below about 300 K a monoclinic phase arises for this composition \cite{Noheda}. We did not incorporate a monoclinic phase in the polydomain model, firstly because the monoclinic phase only becomes observable at low temperatures and secondly its lattice parameters are hard to distinguish from those of the tetragonal phase in our laboratory XRD experiments. In fact the tetragonal lattice parameters are so close together that they are hard to separate and one measures a domain fraction averaged lattice parameter, which varies with substrate induced strain \cite{Noheda}. The adaptive nanodomain model \cite{Steenwelle2012,Jin2003} developed for relaxor materials can also be applied to the PZT system to resolve this ambiguity. In this model the tetragonal domains are so small that the lattice parameters of the tetragonal unit cells cannot be resolved by X-ray diffraction and effective lattice parameters arise that adapt to the misfit strain with a varying $c$-domain fraction \cite{WangPRB2006,WangPRB2007}. More recently it was shown that the adaptive nanodomain state can also arise in the PZT system near the MPB \cite{Rossetti2008,Khachaturyan2010}. Our findings predict that in the clamped film of the PZT(\ce{x_{Ti}}$=0.5$) composition the adaptive polydomain tetragonal phase is energetically the most favorable. Since the model presented in this paper does not impose any restrictions on the domain sizes, a nanodomain sstructure is an allowed solution. Thus this finding supports the applicability of the adaptive, tetragonal nanodomain model for the description of the properties of near MPB compositions of clamped PZT thin films.

The results of the present work are applicable to clamped, epitaxial, (001)-oriented, relatively thick films, in which all strain is relaxed at the deposition temperature. Further development is needed to incorporate the effects of grain boundaries, which are present in many practical films, consisting of closely packed columnar grains which are well oriented in the out-of-plane direction, but often less good or not at all in in-plane directions. The varying elasticity and strength of the mechanical coupling between grains and possible electrical charging of grain boundaries, as well as different in-plane crystallographic orientation of grains are expected to have significant effects on the film properties.

The analytical results of this study allow fairly straightforward comparison with experimental data, needed to test the validity of the model in comparison with earlier models in litereature.

\vspace{2em}
\textbf{5 Conclusions} \newline

The model discussed in this chapter describes the properties of polydomain, (001)-oriented PZT thin films, assuming the presence of three domains in the tetragonal phase. Further the domain walls are assumed not to impose additional boundary conditions on the stresses and polarizations in the domains. The role of the domain walls is only to connect the domains. These assumptions are the main differences with an earlier model in literature. The new assumptions give the system more degrees of freedom to find an energy minimum. It is believed that the present model gives a more realistic description of polydomain epitaxial thin films in which the domain walls can freely move .

The properties of the film were studied analytically as well as numerically. It is found that for the strain values induced by practically used substrates 

a) the tetragonal PZT compositions are always in the polydomain tetragonal $c/b/a$-phase, while the rhombohedral compositions are in the polydomain $r$-phase. The near-MPB PZT($x_{Ti}=0.5)$ composition is found to be in the $c/b/a$-phase. 

b) In the $c/b/a$-phase the stresses in both in-plane directions are equal at finite applied field values and zero at zero field. The elastic energy in the film is therefore zero at zero field and the film is not strained.

c) The analysis allows to decompose the dielectric and piezoelectric properties into components arising from different causes. In the $c/b/a$-phase the dielectric constant of the film, ${\varepsilon }^f_{33}$, is due to domain wall-motion and the rotation of the polarization vector of the in-plane domains, whereas the $c$-domains do not contribute. The piezoelectric constant of the film, $d^f_{33}$, is due to (a) domain wall-motion, (b) the piezoelectric effect of the in-plane domains, while the $c$-domains hardlu contribute to the piezoelectric effect,) and (c) elastic effects depending on the domain fractions. The piezoelectric constant $e^f_{31}$ is not dependent on the domain fractions, but only on the electrostrictive coefficients.

d) In the $r$-phase the polarization rotates under the influence of substrate strain and applied field, whereas the polarization extension is fairly small. Hence the stress energy in the film is reduced by changing the rhombohedral angles of the unit cell. 

\vspace{2em}
\textbf{Acknowledgements}\\
This work was financially supported by NanoNextNL, a micro- and macrotechnology consortium of the Government of the Netherlands and 130 partners and carried out under the project number M62.3.10404 in the framework of the Research Program of the Materials innovation institute (M2i) (www.m2i.nl).

\newpage

\appendix
\textbf{ Appendix A - Estimation of the strain parameter $\mathbf{S_m^0}$ in clamped ferroelectric thin film}
\setcounter{equation}{0}
\renewcommand{\theequation}{A.\arabic{equation}}
\vspace{1em}

The substrate induced thermal (in-plane) strain $S^c_m$ in a film with a strain-free cubic lattice parameter $a_c$ at temperature $T$ (usually room temperature), that is deposited strain-free at deposition temperature $T_d$, can be shown to be given by

\begin{equation}
S^c_m(T)=\left(\frac{a^*_s-a_c}{a_c} \right)_T \approx (\alpha_f-\alpha_s)(T_d-T)
\end{equation}

Here $\alpha_f$ and $\alpha_s$ are the average thermal expansion coefficients of the film and the substrate over the temperature interval $T..T_d$ respectively. $a_c$ is the (cubic) lattice parameter of the strain-free film and the in-plane lattice parameter of the clamped film is equal to the effective substrate parameter at temperature $T$, $a^*_s$. Note that eq.(A1) does not describe epitaxial strain, but only strain due to thermal mismatch. For a film with a non-cubic unit cell (at temperature $T$) one can define an equivalent pseudocubic lattice parameter from the pseudocubic unit cell volume as $a_{pc}=V^{{1}/{3}}_{pc}$, hence for a tetragonal lattice $a_{pc}=\sqrt{a^2_Tc_T}$. Now consider a ferroelectric material. We define $a_0$ as the equivalent cubic lattice parameter of the paraelectric phase at deposition extrapolated to the considered temperature $T$ of the film. We can now write the pseudocubic lattice parameter as $a_{pc}(T)=a_0(T)[(1+S_1)(1+S_2)(1+S_3)]^{1/3}$, where $S_i$ is the total strain in direction \textit{i,} composed of stress induced strain and the stress-free self strains due to the paraelectric-ferroelectric phase transition. For a stress-free unit cell in the tetragonal, ferroelectric phase $S_1=S_2=Q_{12}P^2_s$ , $S_3=Q_{11}P^2_s$ and $a_{pc}\approx a_0(1+(Q_{11}+2Q_{12})P^2_s/3)\ $, hence there is a significant volume change due to the paraelectric-ferroelectric phase transition. For the (high temperature) orthorhombic phase $S_1=S_2=S_3=(Q_{11}+2Q_{12})P^2_s/3$ and thus also $a_{pc}=a_0(1+(Q_{11}+2Q_{12})P^2_s/3). $

The misfit strain as used in the main paper is defined by

\begin{equation}
S^0_m(T)= \left( \frac{a^*_s-a_0}{a_0} \right) _T
\end{equation}

i.e. the misfit is calculated with respect to the equivalent cubic lattice parameter at temperature $T$. To a good approximation one can estimate this misfit strain as (we define $(Q_{11}+2Q_{12})P^2_s/3\equiv Q^*P^2_s$

\begin{equation}
\begin{aligned}
S^0_m(T) &= \left( \frac{a^*_s-a_0}{a_0} \right) _T 
= \left( \frac{a^*_s-a_{pc}}{a_{pc}} \right) _T  (1+Q^*P^2_s)+\left( \frac{a_{pc}-a_0}{a_0}\right)_T \\
&\approx S^c_m(T)+Q^*P^2_s=(\alpha_f-\alpha_s)(T_d-T)+Q^*P^2_s
\end{aligned}
\end{equation}

Here we used $Q^*P^2_s\ll 1$. Hence the misfit strain depends on the thermal expansion coefficients of film and substrate material, as well as its piezoelectric properties due to the paraelectric-ferroelectric phase transition. In table A1 the contribution of the term $Q^*P^2_s$ to the misfit strain is given for several PZT compositions. 

The experimentally accessible strain range by using different substrates is defined by their thermal expansion coefficients. With a substrate thermal expansion range $\alpha _s \approx$ 0-11.7 ppm/K (0 ppm/K for zero thermal expansion glass substrates (known as ULE glass) using oxide nanosheet bufferlayers \cite{Bayraktar2014}, 2.4 for Si, up to 11.5 ppm/K for STO) and $T_d = $ \SI{600}{\celsius} and experimental temperature $T$ \SI{25}{\celsius}s the misfit strain $S^0_m$ is calculated for the different compositions and shown in Table A1. One obtains a fairly narrow accessible misfit strain range of a few 0.1\% around zero for compositions in the range $x_{Ti}=0.4-0.8$, whereas for larger $x_{Ti}$ $S^0_m$ increases rapidly to tensile misfit strains of 1-2\% due to the large value of $Q^*$ for these compositions. The latter is due to the effectively negative thermal expansion coefficients of Ti-rich compositions. For compositions close to the MPB $S^0_m$ is well approximated by $S^c_m$.
The temperature dependence of the misfit strain arises through the terms $S^c_m(T)$ and $P^2_s(T)$.

Finally we note that for a thick film the misfit strain is imposed on the complete film, thus the average in-plane strains in the film must be equal to this value, whereas in a coherently grown epitaxial film, $S^0_m$ is imposed on the individual unit cells of the film. The consequence of this is that a coherently grown film must always be in a single domain phase.
\vspace{1em}

\textbf{Table A1}  
\renewcommand{\arraystretch}{1.5}
\begin{center}
\begin{tabular}{|c|c|c|c|c|c|c|c|c|} \hline
	$x_{Ti}$ & 0.4 & 0.5 & 0.6 & 0.7 & 0.8 & 0.9 & 1.0 \\ \hline 
	$Q^*P^2_s=\frac{1}{3}(Q_{11}+2Q_{12})P^2_s$ & 0.0011 & 0.0004 & 0.0013 & 0.0024 & 0.0033 & 0.0053 & 0.0179 \\ \hline 
	$\alpha_f$$^1$ & 4.8 & 4.4 & 3.5 & 2.2 & 0.4 & -1.8 & -4.3 \\ \hline 
	$S^0_m(RT,on STO)$ & -0.0027 & -0.0037 & -0.0022 & -0.0020 & -0.0011 & 0.0103 & 0.0152 \\ \hline 
	$S^0_m(RT,on Si)$ & 0.0025 & 0.0015 & 0.0031 & 0.0032 & 0.0042 & 0.0155 & 0.0204 \\ \hline 
	$S^0_m(RT,on ULE)$ & 0.0039 & 0.0029 & 0.0044 & 0.0046 & 0.0055 & 0.0169 & 0.0218 \\ \hline 
\end{tabular}
\end{center}

{\footnotesize }$^1$ Average thermal expansion coefficient over the temperature range $T=$ \SI{25}{\celsius} to $T_d=$ \SI{600}{\celsius} is calculated from the dilatation data \cite{Shirane1952} as $\alpha_f=(l_{T_d}-l_T)/(l_T(T_d-T))$. The values for intermediate compositions are obtained from a fitted function $\alpha_f(x_{Ti})=-20.87x^2_{Ti}+13.84x_{Ti}+2.70$.

\newpage 

\appendix
\textbf{ Appendix B - Linearization parameters}
\setcounter{equation}{0}
\renewcommand{\theequation}{B.\arabic{equation}}
\vspace{1em}

The misfit strain and field dependence of the $c$-domain fraction of PZT thin films can be written in a linearized form as \eqref{linSm0}
\begin{align}
\phi(S^0_m,E)=\phi_{0,0}+a_{\phi Sm}S^0_m+a_{\phi E}E
\end{align}

The values of $\phi_{0,0}=\phi (S^0_m=0,E=0)=\frac{Q_{11}+Q_{12}}{Q_{11}-Q_{12}}$ and the derivatives $a_{\phi Sm}=\left( \pdv{\phi}{S^0_m} \right)_{0,0}=\frac{-2}{(Q_{11}-Q_{12})P^2_s}$ and $a_{\phi E,rot}=\left( \pdv {\phi }{E} \right) _{0,0}$ are tabulated in Table B1 for the tetragonal compositions. 

From the table it is seen that with increasing tetragonality of the PZT composition the sensitivity of the domain fraction for changing misfit strain and applied field strongly decreases. Interestingly the effect of strain and applied field (in units of $10^7$ V/m) is very similar

In the same table we give the numerical values of the different contributions to the piezoelectric coefficient as also depicted in fig.\ref{figd33e31}a.

\vspace{1em}
\textbf{Table B1 }
\renewcommand{\arraystretch}{1.5}
\begin{center}
\begin{tabular}{|c|c|c|c|c|c|c|} \hline
	$x_{Ti}$ & 0.5 & 0.6 & 0.7 & 0.8 & 0.9 & 1.0 \\ \hline 
	\multicolumn{7}{|c|}{Domain wall motion} \\ \hline 
	$\phi_{0,0}$ & 0.35 & 0.47 & 0.52 & 0.54 & 0.54 & 0.55 \\ \hline 
	$a_{\phi Sm}$ & -56 & -55 & -57 & -39 & -33 & -9 \\ \hline 
	$a_{\phi E,rot}$($E$ in units of 100kV/cm) & 0.054 & 0.051 & 0.054 & 0.029 & 0.005 & 0.003 \\ \hline 
	\multicolumn{7}{|c|}{Piezoelectric effect} \\ \hline 
	$d^{DW}_{33}\ $(pm/V) & 286 & 236 & 219 & 179 & 163 & 114 \\ \hline 
	$d^{PE}_{33}$ (pm/V) & -91 & -33 & -12 & -9 & -8 & -13 \\ \hline 
	$d^{elas}_{33}$ (pm/V) & -104 & -88 & -89 & -70 & -61 & -31 \\ \hline 
\end{tabular}
\end{center}

\pagebreak
\setcounter{equation}{0}
\setcounter{figure}{0}
\setcounter{table}{0}
\setcounter{page}{1}
\renewcommand{\theequation}{S\arabic{equation}}
\renewcommand{\thefigure}{S\arabic{figure}}




	\noindent \textbf{Supplementary Information}
	\vspace{1em}
	\begin{center}
		\Large 
		\textbf{Modelling functional properties of ferroelectric oxide thin films}
		\textbf{with a three-domain structure}
	\end{center}	
	\normalsize
	\vspace{2em}
	
	\noindent \textbf{E.P. Houwman$^1$, K. Vergeer$^{1,2}$, G. Koster$^1$ and G. Rijnders$^1$ }
	
	\vspace{2em}
	
	\noindent $^1$ Inorganic Material Science, MESA+ Institute of Nanotechnology, University of Twente, \\ Enschede, The Netherlands
	
	\noindent $^2$ Materials innovation institute (M2i), Delft, The Netherlands
	
	\noindent Corresponding author: e.p.houwman@utwente.nl
	\vspace{2em}
	
	In the main paper mainly results are given for the (near) zero-field polydomain $c/a/b$ phase and the non-zero field $c/bc^*/ac^*$ phase. Here we give the results for the other phases, appearing in the phase diagrams. Several results for some of these phases have been given before \cite{S_Koukhar2001,S_Perstev2003,S_Kukhar2006}. These are reproduced here and some more analytical results are given for future reference. We will indicate the property or parameter $p$ of the clamped thin film in a certain phase (for example $c$ or $c/a/b$) with a superindex $y$, whereas the property of a parameter of a given domain $x$ in phase $y$ is given by a subindex $x$, eventually followed by a second (and third) subindex $i$ ($j$) for the directional component $p^y_{xij}$ of that parameter.
	
	We remind the reader here that the only temperature dependent parameter in the Landau-Devonshire formulation of the Gibb's energy of the \ce{PbZr_{1-x}Ti_xO_3} ferroelectric \cite{S_Haunseries} is the electric stiffness coefficient given by a linear temperature dependence, $\alpha_1=(T-\theta)/2\varepsilon_0C$, with $C$ the Curie-Weiss constant, $\theta $ the Curie-Weiss temperature and $\varepsilon_0$ the permittivity of free space. For a second-order paraelectric-ferroelectric phase transition ${\alpha }_1$ vanishes at the transition temperature $T_C=\theta $, whereas for a first-order transition ${\alpha }_1$ is finite at the transition temperature $T_C>\theta $.
	
	\vspace{2em}
	\textbf{1 Single domain \textbf{\textit{c}}-phase}
	\vspace{1em}
	
	Symmetric in-plane clamping causes compressive in plane stress, so that only $c$-domains exist if $S^0_m<Q_{12}(P^c_s)^2$. The saturation polarization $P^c_s$ in the latter relation is equal to that of a stress-free bulk sample, given by \cite{S_Haunseries}
	$P^{blk}_s=\left[ -\frac{\alpha_{11}}{3\alpha_{111}}+\left( \left( \frac{\alpha_{11}}{3\alpha_{111}}\right)^2-\frac{\alpha_1}{3\alpha_{111}}\right)^{1/2} \right]^{1/2}$ for tetragonal compositions with a second-order paraelectric-ferroelectric phase transition, i.e. for \ce{x_{Ti}}$\le 0.717$. For stress-free bulk samples with compositions in the range $0.717<\ce{x_{Ti}}\le 1$ the paraelectric-ferroelectric phase transition is first order. The temperature dependence of the polarization $P^c(T)$ of a stress free sample is in that case described by ${P^c(T)}^2=\Psi(T){()P^c_C)}^2$, with $P^c_C$ the polarization at the critical (phase transition) temperature $T_C$. $P^c_C$is given by the relation $\alpha_{11}=-(T_C-\theta)/\varepsilon_0C(P^c_C)^2$ or $\alpha_{111}=(T_C-\theta)/2\varepsilon_0C(P^c_C)^4$ and the scaling function by $\Psi(T)=(2/3){1+[1-3(T-\theta)/4(T_C-\theta)]^{1/2}}$ \cite{S_Haunseries}.
	
	Now we return to the case of a compressively stressed thin film. The in-plane stress (for zero and finite applied field) in this mono-domain state is \cite{S_Zembilgotov1989}.
	
	\begin{equation}
	\begin{aligned}
	\sigma =\sigma_1&=\sigma_2=\frac{S^0_m-Q_{12}{P^c}^2}{s_{11}+s_{12}} \\ 
	\sigma_3=\sigma_4&=\sigma_5=\sigma_6=0 
	\end{aligned}
	\end{equation}
	
	Here $P^c$ is the polarization of the $c$-phase and is equal to the measured polarization $P_m$ in a parallel plate capacitor geometry. The average strains in the film are equal to those in the individual unit cells and are given by
	
	\begin{equation}
	\begin{aligned}
	S_1=S_2&=S^0_m \\
	S_3=\frac{2s_{12}S^0_m}{s_{11}+s_{12}}&+\left(Q_{11}-\frac{2s_{12}Q_{12}}{s_{11}+s_{12}}\right){P^c}^2
	\end{aligned}
	\end{equation}
	
	Substituting the strains in the general free energy expression (7) with ${\phi }_c=1$ results into
	
	\begin{equation}
	\label{Fc}
	F^c=\frac{{\left(S^0_m\right)}^2}{s_{11}+s_{12}}+{\alpha }^*_1{P^c}^2+{\alpha }^*_{11}{P^c}^4+{\alpha }_{111}{P^c}^6-EP^c
	\end{equation}
	
	with ${\alpha }^*_1={\alpha }_1-\frac{2Q_{12}S^0_m}{s_{11}+s_{12}}$ and ${\alpha }^*_{11}={\alpha }_{11}+\frac{Q^2_{12}}{s_{11}+s_{12}}$ (NB the notation is different from [2]). From this follows that the zero-field polarization value in the compressively strained monodomain $c$-phase, $P^c_s\ $, is significantly enhanced over that of the unstrained monodomain $c$-phase, $P^{blk}_s$, as
	
	\begin{equation}
	\label{Pc2(0)}
	P^{c2}(0)=P^{c2}=-\frac{{\alpha }^*_{11}}{3{\alpha }_{111}}+{\left({\left(\frac{{\alpha }^*_{11}}{3{\alpha }_{111}}\right)}^2-\frac{{\alpha }^*_1}{3{\alpha }_{111}}\right)}^{\frac{1}{2}}
	\end{equation}
	
	\noindent Following the same procedure as in [4] for the compositions with a first-order phase transition one finds the critical temperature $T^*_C$ and polarization $P^{c*}_C$ at this temperature of the clamped film. Solving these parameters from these relations one has $(T^*_C-\theta)=(\varepsilon_0\alpha^{*2}_{11}C)/2\alpha_{111}$ and $(P^{c*}_C)^2=-\alpha^*_{11}/2\alpha_{111}$.
	Following again Haun's procedure one has finally for the temperature dependence of the polarization of the clamped film $P^{c2}=\Psi^*(T)(P^{c*}_C)^2$ with the scaling function ${\mathit{\Psi}}^*(T)=(2/3) \left( 1+\left[1-3(T-\theta)/4(T^*_C-\theta)\right]^{1/2} \right) $.
	
	The relation defining the polarization loop of the second order ferroelectric follows from minimizing \eqref{Fc} with respect to $P^c$
	
	\begin{equation}
	\label{2a1Pc}
	{2\alpha }^*_1P^c+4{\alpha }^*_{11}{P^c}^3+6{\alpha }_{111}{P^c}^5=E
	\end{equation}
	
	The derivation of the loop equation for the first order material is slightly more complicated. Again following the procedure of \cite{S_Haunseries} but now for finite field one finds
	$\alpha^*_{11}=(T^*_C-\theta)/(\varepsilon_0P^{c*2}_CC)+5E/2P^{c*3}_C$ and $\alpha_{111}=(T^*_C-\theta)/\left(2\varepsilon_0P^{c*4}_CC\right)-3E/2$, from which $P^{c*}_C(E)$ and $T^*_C(E)$ are found. The polarization loop is then calculated as $P^{c2}(E)=\Psi^*(T,E)P^{c*2}_C(E)$
	
	The ferroelectric-paraelectric transition temperature of the mono $c$-domain film, $T^c_C$, is found from the condition ${\alpha }^*_1=0$ as
	
	\begin{equation}
	T^c_C=2{\varepsilon }_0C\frac{2Q_{12}S^0_m}{s_{11}+s_{12}}+\theta
	\end{equation}
	
	From \eqref{2a1Pc} follows the dielectric constant in the $c$-phase as
	
	\begin{equation}
	{\left(\frac{\partial P^c}{\partial E}\right)}_E={\left({2\alpha }^*_1+12{\alpha }^*_{11}{P^c}^2+30{\alpha }_{111}{P^c}^4\right)}^{-1}={\varepsilon }_0{\varepsilon }^c_{33}\left(E\right)
	\end{equation}
	
	which in form is equal to the bulk result but with stiffness coefficients that are modified by the clamping.
	
	The piezoelectric constant of the $c$-domain is given by
	
	\begin{equation}
	d^c_{33}(E)=\left(\frac{\partial S_3}{\partial E}\right)^\sigma =\left(Q_{11}-\frac{2s_{12}Q_{12}}{s_{11}+s_{12}}\right) 2P^c \left(\frac{\partial P^c}{\partial E}\right)^E
	\end{equation} 
	
	Defining $d^*_{33}=2Q_{11}P^c{\left(\frac{\partial P^c}{\partial E}\right)}_E$ and $d^*_{31}=2Q_{12}P^c{\left(\frac{\partial P^c}{\partial E}\right)}_E$ this can be rewritten in the often used expression for a clamped thin film
	
	\begin{equation}
	\label{dc33}
	d^c_{33}=d^*_{33}-\frac{2s_{12}}{s_{11}+s_{12}}d^*_{31}
	\end{equation} 
	
	with the distinction that ${\left(\frac{\partial P^c}{\partial E}\right)}_E$ and $P^c$, defining the piezoelectric parameters $d^*_{3i}$ , are given by the relations \eqref{2a1Pc} and \eqref{Pc2(0)} for the clamped film, which are not equal to those of the bulk material. Thus although \eqref{dc33} is often used by experimentalists, the use of bulk values for the piezoelectric coefficients in \eqref{dc33} is not correct. Further on we will see that for other domain configurations the difference with the usually used expression $d^{clamped}_{33}=d^{blk}_{33}-\frac{2s_{12}}{s_{11}+s_{12}}d^{blk}_{31}$ is even larger.
	The other piezoelectric coefficients $d_{c3i}$ are zero. 
	
	The non-zero $e^c_{3i}$ coefficients are
	
	\begin{equation}
	e^c_{31}=e^c_{32}=-{\left(\frac{\partial {\sigma }_1}{\partial E}\right)}^S=\frac{Q_{12}}{s_{11}+s_{12}}2P^c{\left(\frac{\partial P^c}{\partial E}\right)}^S
	\end{equation} 
	
	From \eqref{2a1Pc} it follows that ${\left(\frac{\partial P^c}{\partial E}\right)}^E={\left(\frac{\partial P^c}{\partial E}\right)}^S$ hence $e^c_{31}={d^c_{31}}/{\left(s_{11}+s_{12}\right)}$.
	
	(S6-10) also apply for the first-order transition compositions, using the appropriate $P^c(E)$ dependence.
	
	\vspace{2em} 
	\textbf{2 Polydomain \textbf{\textit{a/b}}-phase} \newline

	The polydomain $a/b$-phase (an alternative name used is the $a_1/a_2$-phase \cite{S_Koukhar2001}) arises under tensile stress, when all \textit{c}-domains are pulled into the film plane. The polarization vectors are aligned along the in-plane $[100]$ and $[010]$ vectors. By symmetry the domain fractions of \textit{a} and \textit{b} domains must be equal, $\phi _a=\phi_b=0.5$, unless asymmetric strain conditions are applied to the film \cite{S_Zembilgotov1989,S_Qiu2010}. We expect that for thick films any anisotropy in the in-plane lattice constants of the substrate is likely to be removed during growth by strain relaxation induced by defects. To achieve anisotropy in two-dimensional films at room temperature requires anisotropic thermal expansion in the in-plane directions of the substrate. Anisotropic strain conditions are present in thin films stressed in one direction, for example in bent cantilever structures. Anisotropic strain conditions also apply in very narrow structures, such as transmission electron microscope samples. For very thin films, that are grown cube-on cube on perovskite substrates, thus without strain relaxation by defects, anisotropic strain conditions by the substrate results may also be applicable.
	
	Here we assume symmetric in-plane clamping. From this follows that $\sigma_{a1}=\sigma_{b2}$, $\sigma_{a2}=\sigma_{b1}$. Further the macroscopic boundary conditions $<\sigma_3> =<\sigma_4> =<\sigma_5> =0$ result in the conditions $\sigma_{a3}=-\sigma_{b3}$, $\sigma_{a4}=-\sigma_{b4}$, $\sigma_{a5}=-\sigma_{b5}$ From the condition $<S_6>=0$, follows $\sigma_{a6}=-\sigma_{b6}\ $. Assuming the \textit{strict electrical boundary condition} on the domain walls --thus no polarization rotation- to be applicable gives $P_{a1}=P_{b2}=P^{ab}$ and no polarization rotation out-of-plane. From the equations of state for the in-plane strain conditions one obtains
	
	\begin{equation}
	2S^0_m=\left(s_{11}+s_{12}\right)\left({\sigma }_{a1}+{\sigma }_{a2}\right)+\left(Q_{11}+Q_{12}\right)P^{ab2}
	\end{equation}
	
	Hence 
	
	\begin{equation}
	\label{eqsigma-a2}
	\sigma_{a2}=\frac{2S^0_m-\left({Q_{11}+Q}_{12}\right)P^{ab2}-\left(s_{11}+s_{12}\right){\sigma }_{a1}}{s_{11}+s_{12}}\equiv A-{\sigma }_{a1}
	\end{equation} 
	
	From the \textit{strict mechanical domain wall boundary conditions} it follows that ${\sigma }_{a1}={\sigma }_{b1}={\sigma }_{a2}={\sigma }_{b2}=\sigma $ and ${\sigma }_{a3}={\sigma }_{b3}={\sigma }_{a4}={\sigma }_{b4}={\sigma }_{a5}={\sigma }_{b5}={\sigma }_{a6}={\sigma }_{b6}=0$. The latter conditions also follow from minimizing the free energy with respect to ${\sigma }_{a3}$, ${\sigma }_{a4}$, ${\sigma }_{a5}$ and ${\sigma }_{a6}$ respectively. Similarly the first condition for the in-plane stresses can be obtained from \eqref{eqsigma-a2} without invoking the \textit{strict mechanical domain wall boundary conditions}:
	
	\begin{equation}
	\begin{aligned}
	\sigma &=\sigma_{a1}=\sigma_{b1}=\sigma_{a2}=\sigma_{b2}=A/2=\frac{S^0_m-\frac{1}{2}\left({Q_{11}+Q}_{12}\right)P^{ab2}}{s_{11}+s_{12}} \\
	\sigma_{a3} &=\sigma_{b3}=\sigma_{a4}=\sigma_{b4}=\sigma_{a5}=\sigma_{b5}=\sigma _{a6}=\sigma_{b6}=0
	\end{aligned}
	\end{equation} 
	
	\noindent The strains are then given by
	
	\begin{equation}
	\begin{aligned}
	S_{a1}=S_{b2}&=S^0_m+\frac{1}{2}\left({Q_{11}-Q}_{12}\right)P^{ab2} \\
	S_{a2}=S_{b1}&=S^0_m-\frac{1}{2}\left({Q_{11}-Q}_{12}\right)P^{ab2} \\
	S_{a3}=S_{b3}&=Q_{12}P^{ab2} \\
	S_{a4}=S_{b4}&=S_{a5}=S_{b5}=S_{a6}=S_{b6}=0
	\end{aligned}
	\end{equation} 
	
	\noindent Substituting the strain in the free energy expression results into
	
	\begin{equation}
	F^{ab}=\frac{{S^0_m}^2}{s_{11}+s_{12}}+\alpha^{**}_1P^{ab2}+\alpha^{**}_{11}P^{ab4}+\alpha_{111}P^{ab6}
	\end{equation} 
	
	with ${\alpha }^{**}_1={\alpha }_1-\frac{\left({Q_{11}+Q}_{12}\right)S^0_m}{s_{11}+s_{12}}$ and ${\alpha }^{**}_{11}={\alpha }_{11}+\frac{{\left({Q_{11}+Q}_{12}\right)}^2}{4\left(s_{11}+s_{12}\right)}$. 
	
	\begin{sloppypar}
		\noindent From this follows the zero-field polarization for the second-order phase transition materials in both domains as $P_{a1}=P_{b2}\equiv P^{ab}(E=0)$ given by an equation analogous to \eqref{Pc2(0)}, but with $\alpha^{**}_1$ and $\alpha^{**}_{11}$ replacing $\alpha^*_1$ and $\alpha^*_{11}$, respectively. For the first-order phase transition materials one has, analogous to the previous section, $(T^{**}_C-\theta)=(\varepsilon _0\alpha^{**2}_{11}C)/2\alpha_{111}$ and $(P^{ab*}_C)^2=-\alpha^{**}_{11}/2\alpha_{111}$ , and $(P^{ab})^2=\Psi^{**}(T)(P^{ab*}_C)^2$ with $\Psi^{**}(T)=(2/3)\left(1+[1-3(T-\theta)/4(T^{**}_C-\theta)]^{1/2}\right)$. 
	\end{sloppypar}
	
	Since it is in the used approximation assumed that there is no coupling of the polarization with the applied field in the 3-direction, the polarization is field independent and consequently all stress and strain components are field independent as well.
	
	The $a/b$-phase state changes into the polydomain $c/a/b$ phase when the stress becomes zero at $S^0_m=\frac{1}{2}({Q_{11}+Q}_{12})P^{ab2}$ and the in plane strains are equal to the bulk strains, and also then $P^{ab}(0)=P^{blk}_s$.
	
	The ferroelectric-paraelectric transition temperature is found again from the condition ${\alpha }^{**}_1=0$ as
	
	\begin{equation}
	T^{ab}_C=2\varepsilon_0C\frac{({Q_{11}+Q}_{12})S^0_m}{s_{11}+s_{12}}+\theta
	\end{equation}
	
	Since the polarization is independent of the applied field in the 3-direction, the net dielectric and piezoelectric parameters obey
	
	\begin{equation}
	\varepsilon^{ab}_{3i}=d^{ab}_{3i}=e^{ab}_{3i}=0
	\end{equation} .
	
	\vspace{2em}
	\textbf{3 Polydomain \textbf{\textit{ac$^*$/bc$^*$}}-phase: the \textbf{\textit{a/b}}-phase in applied electrical field}
	\vspace{1em}
	
	Under the influence of an external $E_3$ field one expects that the polarization in the in-plane oriented domains rotates in the out-of-plane direction, creating a small $P_{x3}$ component to the polarization in the $a$ and $b$ domains of the $a/b$ phase, creating a new ${ac^*}/{bc^*}$ phase. The free energy of the ${ac^*}/{bc^*}$ domain structure\textbf{ }therefore contains the terms due to the ${P_{a1}=P_{b2}=P}_1$ and ${P_{a3}=P_{b3}=P}_3$ components of the polarization and can be written as (here we drop for convenience the superindex $ac^*bc^*$ to the polarization components $P_1$ and $P_3$)
	
	\begin{equation}
	\label{eqFabc*}
	F^{ac*bc*}=\alpha_1(P^2_1+P^2_3)+\alpha_{11}(P^4_1+P^4_3)+\alpha_{111}(P^6_1+P^6_3)+\alpha_{12}P^2_1P^2_3+\alpha_{112}(P^4_1P^2_3+P^4_3P^2_1) 
	+(s_{11}+s_{12})\sigma^2-EP_3
	\end{equation}
	
	Here we have already used the result that all stress components are equal zero, except $\sigma_{a1}=\sigma_{b2}=\sigma$ and $\sigma_{a2}=\sigma_{b1}=\sigma$, as follows from considering the boundary conditions.
	
	The strains are now a function of both polarization components
	
	\begin{equation}
	\begin{aligned}
	\label{eqstrains}
	S_{a1}=S_{b2}&=(s_{11}+s_{12})\sigma +Q_{11}P^2_1+Q_{12}P^2_3 \\
	S_{a2}=S_{b1}&=(s_{11}+s_{12})\sigma +Q_{12}P^2_1+Q_{12}P^2_3 \\ 
	S_{a3}=S_{b3}&=2s_{12}\sigma +Q_{11}P^2_1+Q_{11}P^2_3 \\ 
	S_{a4}=S_{b5}&=Q_{44}P_1P_3 \\
	S_{a5}=S_{b4}&=S_{a6}=S_{b6}=0
	\end{aligned}
	\end{equation}
	
	Substituting these in the boundary condition $S^0_m=<S_1> =\frac{1}{2}(S_{a1}+S_{b1})$ it follows that
	
	\begin{equation}
	\label{eqstress}
	\sigma =\frac{S^0_m-Q_{12}P^2_3-\frac{1}{2}(Q_{11}+Q_{12})P^2_1}{s_{11}+s_{12}}
	\end{equation}
	
	Substitution into the free energy expression reduces the number of variables to two: $F^{ac*bc*}(P_1,P_3)$. Minimization of $F^{ac*bc*}$ with respect to $P_3$ in the limit of small fields, so that $\left|P_3\right|\ll \left|P_1\right|$, and taking $P_1\approx P_{10}\equiv P_1(S^0_m,E=0)$, as defined above for the $a/b$ phase, one obtains a linear field dependence of the out-of-plane polarization, $P_3=\varepsilon_0\varepsilon^{ac^*bc^*}_{33}E$, with
	
	\begin{equation}
	\label{eqdP3dE}
	\varepsilon_0\varepsilon^{ac^*bc^*}_{33}=\left(\frac{\partial P_3}{\partial E}\right)_0=\frac{1}{2\left(\alpha_1+\alpha_{12}P^2_{10}+\alpha_{112}P^{4}_{10}-\frac{2S^0_mQ_{12}}{s_{11}+s_{12}}+\frac{Q_{12}(Q_{11}+Q_{12})}{s_{11}+s_{12}}P^2_{10}\right)}
	\end{equation}
	Thus the susceptibility  of the $a/b$ phase in a field (in the 3-direction) is field independent and the film shows the linear behaviour of a paraelectric material.
	The angle $\theta $ of the polarization vector with the film plane is then for small fields given by
	
	\begin{equation}
	\label{eqtan}
	tan\theta =\frac{P_3}{P_1}\approx E\frac{\varepsilon_0\varepsilon^{ac^*bc^*}_{33}}{P_{10}}
	\end{equation}
	
	Minimization of $F^{ac*bc*}$ with respect to $P_1$ in the limit of small fields, where $P_3=0$, gives the same field independent polarization $P_{10}\mathrm{\ }$as for the $a/b$ domain structure, and consequently ${\left({\partial P_1}/{\partial E}\right)}_0=0$, hence $\varepsilon^{ac^*bc^*}_{31}=0$.
	From (\eqref{eqstress}) it follows with the above results that the derivative $\left({\partial \sigma }/{\partial E}\right)_0=0$ \textit{in lowest order}, therefore also the piezoelectric constants (for $E=0$) $d^{ac^*bc^*}_{33}=\left({\partial {<S_3>}}/{\partial {E}}\right)_0=0$ and  $e^{ac^*bc^*}_{31}=-\left({\partial \sigma}/{\partial{E}}\right)_0=0$.
	
	Since $P_1\approx P_{10}$ at low fields and $P_3=\varepsilon_0\varepsilon^{ac^*bc^*}_{33}E$ the in-plane stress decreases in second order quadratically with the applied field as
	
	\begin{equation}
	\label{eqstress(E)}
	\sigma (E)
	=\frac{S^0_m-Q_{12}(\varepsilon_0\varepsilon^{ac^*bc^*}_{33}E)^2-\frac{1}{2}(Q_{11}+Q_{12})P^2_{10}}{s_{11}+s_{12}}
	\end{equation}
	The field dependent term in \eqref{eqstress(E)} only becomes significant compared to the other terms close to the phase boundary with the $c/a/b$ or $r$-phase, where $\varepsilon^{ac^*bc^*}_{33}$ becomes large, and where $\sigma(0)$ is close to zero (Fig.7b in the main paper), thus only there the stress is field dependent and even becomes tensile for large enough fields (remember that $Q_{12}<0$). This is observed in fig.6d in the main paper for the PZT(\ce{x_{Ti}}$=0.4$) composition for which an $ac^*/bc^*$ to $r$-phase transition is observed at large enough fields at strain values close to the phase boundary. Similarly an $ac^*/bc^*$ to $c/ac^*/bc^*$-phase transition is observed for the PZT(\ce{x_{Ti}}$=0.4$) composition.
	
	From \eqref{eqstress(E)} it is seen that the field dependence of the lattice parameters is largely due to the polarization rotation out-of-plane. The polarization rotation also causes shear strains, which promotes the $ac^*/bc^*$ to the $r$- or the $c/ac^*/bc^*$ phase transition. For a polydomain film one expects from symmetry arguments no net shear in both out-of-plane directions, in contrast with a single domain film. This gives rise to an extra set of macroscopic boundary conditions, which was not considered previously,
	
	\begin{equation}
	\label{eqshearstrains}
	<S_4>=<S_5>=0
	\end{equation}
	
	These conditions are met if equal fractions of $a$, respectively $b$ domains shear in opposite direction, thus these fractions have opposite in-plane polarization vector orientations. Further one could expect that the shearing has consequences for the amount of polarization rotation possible in a thick film, since the shearing causes traction forces on adjacent domains that do not shear in the same direction. The reaction forces oppose the shearing and thus the polarization rotation. One would expect that this rotation frustration effect should be less for materials with smaller $Q_{44}$. Since the shearing is expected to be small for moderate applied fields we neglect this effect. 
	The interaction forces between domains impose microscopic boundary conditions on the domain wall. As discussed in the paper we do not impose microscopic boundary conditions on the domain walls. The consequence of the macroscopic boundary conditions \eqref{eqshearstrains} is therefore only that there are equal fractions of $a$ and $b$ domains, shearing in opposite directions.
	
	\vspace{2em}
	\textbf{4 Polydomain \textbf{\textit{c/ac$^*$/bc$^*$}}-phase: the \textbf{\textit{c/a/b}}- phase in an electrical field}
	\vspace{1em}
	
	The free energy is now a function of the $c$-domain fraction and the polarization rotation in the $a$- (and $b$-) domain. Eq.(7) can therefore be explicitly written as (again we drop for convenience the superindex to the polarization components)
	\begin{sloppypar}
	\begin{equation}
	\begin{aligned}
	\label{eqFcbac*}
	&F^{cac*bc*}=\qquad \qquad \qquad \\
	&\phi\left[\alpha_1P^2_{c3}+\alpha_{11}P^4_{c3}+\alpha_{111}P^6_{c3}-EP_{c3}\right] \\ 
	&+(1-\phi) \left[ \alpha_1(P^2_{a1}+P^2_{a3})+\alpha_{11}(P^4_{a1}+P^4_{a3})+\alpha_{111}(P^6_{a1}+P^6_{a3})+\alpha_{12}P^2_{a1}P^2_{a3}
	+\alpha_{112}(P^4_{a1}P^2_{a3}+P^4_{a3}P^2_{a1})-EP_{a3}  \right] \\
	&+(s_{11}+s_{12})\sigma^2
	\end{aligned}
	\end{equation}
	\end{sloppypar}
	where we have already made the steps of minimization with respect to most stress components and of stress cancellations arising from the boundary conditions, similar as leading to Eq.\eqref{eqFabc*}.
	
	The strains are )
	\begin{equation}
	\begin{aligned}
	S_{a1}=S_{b2}&=(s_{11}+s_{12})\sigma +Q_{11}P^2_{a1}+Q_{12}P^2_{a3} \\ 
	S_{a2}=S_{b1}&=(s_{11}+s_{12})\sigma +Q_{12}P^2_{a1}+Q_{12}P^2_{a3} \\ 
	S_{c1}=S_{c2}&=(s_{11}+s_{12})\sigma +Q_{12}P^2_{c3} \\ 
	S_{a3}=S_{b3}&=2s_{12}\sigma +Q_{12}P^2_{a1}+Q_{11}P^2_{a3} \\ 
	S_{c3}&=2s_{12}\sigma +Q_{11}P^2_{c3} \\
	S_{a5}=S_{b4}&=Q_{44}P_1P_3 \\ 
	S_{c4}=S_{c5}&=S_{a4}=S_{b5}=S_{a6}=S_{b6}=0
	\end{aligned}
	\end{equation} 
	
	From the clamping conditions the stress is obtained as
	\begin{equation}
	\label{eqSigma-cabc*}
	\sigma =\frac{S^0_m-\phi Q_{12}P^2_{c3}-(1-\phi)Q_{12}P^2_{a3}-(1-\phi)(Q_{11}+Q_{12})P^2_{a1}/{2}}{s_{11}+s_{12}}
	\end{equation} 
	
	Substitution of \eqref{eqSigma-cabc*} into \eqref{eqFcbac*}, minimizing the free energy with respect to $P_{a3}$ and expanding $P_{a3}$ to the lowest order in $E$, results into 
	
	\begin{equation}
	\label{Pa3}
	P_{a3}=\frac{E}{2(\alpha_1+\alpha_{12}P^2_s+\alpha_{112}P^4_s)}
	\end{equation}
	
	Here we made use of the equality $P_{a1}(0)=P_{c3}(0)=P_s=P^{blk}_s$ at zero field and the zero-field domain fraction $\phi_0$. $P^{blk}_s$ has been defined previously. Thus the out-of-plane polarization component of the in-plane oriented $a$ and $b$ domains adds a `paraelectric' contribution to the out-of-plane polarization 
	\begin{equation} 
	\varepsilon_0\varepsilon_{a33}=\left(\pdv{P_{a3}}{E}\right)_0 =\frac{1}{2\left(\alpha_{1}+\alpha_{12}P^{2}_0+\alpha_{112}P^{4}_0\right)}
	\end{equation}
	
	In principle also expressions for $\left({\partial P_{a1}}/{\partial E}\right)_0=\varepsilon _0\varepsilon_{a31}$ and $\left({\partial P_{c3}}/{\partial E}\right)_0=\varepsilon_0\varepsilon_{c33}$ for the dielectric constant of the long-axis dielectric constant under a field in the 3-direction for respectively the $a$ and $c$ domain can be derived, but that results in awkwardly complicated expressions and is not pursued further here. 
	
	\vspace{2em}
	\textbf{5 The polydomain \textbf{\textit{r$_1$/r$_2$/r$_3$/r$_4$}}-phase (or \textbf{\textit{r}}-phase) }
	\vspace{1em}
	
	The polydomain $r_1/r_2/r_3/r_4$-phase (or short or $r$-phase) is described by a polarization vector that can rotate in the {110} planes of four different domains. 
	From symmetry follows that all domain fractions are equal to $\phi_x=1/4$ and $\sigma_{a4}=\sigma_{b4}=\sigma_{c4}=\sigma_{d4}=\sigma_{a5}=\sigma_{b5}=\sigma_{c5}=\sigma_{d5}\equiv \sigma_4$. The macroscopic conditions $<\sigma_4> =<\sigma_5>=0$ then make $\sigma_{x4}=\sigma_{x5}=0$. Symmetry also requires that $\sigma_{a1}=\sigma_{b1}=\sigma_{c1}=\sigma_{d1}=\sigma_{a2}=\sigma_{b2}=\sigma_{c2}=\sigma_{d2}\equiv \sigma $. From $<\sigma_3>=0$ it follows that $\sum{\sigma_{x3}}=0$ and from $<S_6>=0$ that $\sum{\sigma_{x6}}=0$. Using the symmetry relations $S_{a1}=S_{d1}$, $S_{a2}=S_{b2}$, etc. it is seen that $\sigma_{a3}=\sigma_{b3}=\sigma_{c3}=\sigma_{d3}=0$
	
	The clamping to the substrate, $S_m=<S_1>=<S_2>=S_{x1}=S_{x2}$ (where $x=a,b,c,d$) results in
	\begin{equation}
	\label{eqstress-r}
	\sigma =\sigma_1=\sigma_2=\frac{S_m^0-(Q_{11}+Q_{12})P^2_1-Q_{12}P^2_3}{s_{11}+s_{12}}
	\end{equation}
	and the free energy is given by
	\begin{equation}
	\label{eqFr}
	\begin{aligned}
	F^r=&\alpha_1\left(2P^2_1+P^2_3\right)+\alpha_{11}\left(2P^4_1+P^4_3\right)+\alpha_{111}\left(2P^6_1+P^6_3\right) \\ &+\alpha_{12}\left(P^4_1+2P^2_1P^2_3\right)+\alpha_{112}\left(2P^6_1+2P^4_3P^2_1+2P^4_1P^2_3\right)+\alpha_{123}P^4_1P^2_3+(s_{11}+s_{12})\sigma^2-EP_3
	\end{aligned}
	\end{equation}
	
	\eqref{eqstress-r} shows that tensile stress will pull the polarization in the film plane as expected and compressive stress will rotate it out-of-plane, but both processes will increase the polarization energy term in the free energy. The minimum of the free energy (for $E=0$) determines the equilibrium polarization angle of the clamped rhombohedral film.
	
	The strains are now given by
	\begin{equation}
	\label{eqstrains-r}
	\begin{aligned}
	S_{a1}=S_{a2}=S_{b1}=S_{b2}=&S_{c1}=S_{c2}=S_{d1}=S_{d2}=S^0_m \\ 
	S_{a3}=S_{b3}=S_{c3}=S_{d3}=\frac{2s_{12}}{s_{11}+s_{12}}S^0_m+&2\left[Q_{12}-\frac{s_{12}(Q_{11}+Q_{12})}{s_{11}+s_{12}}\right]P^2_1+\left[Q_{11}-\frac{2s_{12}Q_{12}}{s_{11}}+s_{12}\right]P^2_3  \\
	S_{a4}=S_{a5}=S_{b4}=S_{b5}=&S_{c4}=S_{c5}=S_{d4}=S_{d6}=Q_{44}P_1P_3 \\ 
	S_{a6}=S_{b6}=&S_{c6}=S_{d6}=Q_{44}P^2_1
	\end{aligned}
	\end{equation}
	
	We were not able to derive simple expressions for piezoelectric or dielectric constants.



\begin{thebibliography} {99}
\bibitem{remarkpoling}  
	The poling is not a necessary condition for the following derivation. It can be shown that the last equality in eq.\eqref{eqFcba} holds in the presence of $a/b$ domain structures, provided the volume of $a$ domains is equal to that of $b$ domains.
\end{thebibliography}
\end{document}